\documentclass{aastex631}

\usepackage{subfigure}
\usepackage{tabularx}
\usepackage{amsmath}
\usepackage{amssymb}

\shorttitle{CaT metallicities of RRLs}
\shortauthors{Kunder et al.}

\begin{document}

\title{The Galactic Bulge exploration III.: Calcium Triplet Metallicities for RR Lyrae Stars }

\author[0000-0002-2808-1370]{Andrea Kunder}
\affiliation{Saint Martin's University, 5000 Abbey Way SE, Lacey, WA, 98503, USA}

\author[0000-0001-5497-5805]{Zdenek Prudil}
\affiliation{European Southern Observatory, Karl-Schwarzschild-Strasse 2, 85748 Garching bei M\"{u}nchen, Germany}

\author[0009-0009-4825-429X]{Claire Skaggs}
\affiliation{Saint Martin's University, 5000 Abbey Way SE, Lacey, WA, 98503, USA}

\author[0000-0001-5761-6779]{Henrique Reggiani}
\affiliation{Gemini Observatory/NSF’s NOIRLab, Casilla 603, La Serena, Chile}

\author[0000-0001-5825-4431]{David M. Nataf}
\affiliation{Department of Physics \& Astronomy, The Johns Hopkins University, Baltimore, MD 21218, USA}

\author[0000-0002-9074-0306]{Joanne Hughes}
\affiliation{Physics Department Seattle University, 901 12th Ave., Seattle, WA 98122, USA}

\author[0000-0001-6914-7797]{Kevin R. Covey}
\affiliation{Department of Physics \& Astronomy, Western Washington University, MS-9164, 516 High St., Bellingham, WA, 98225}

\author[0000-0002-3723-6362]{Kathryn Devine}
\affiliation{The College of Idaho, 2112 Cleveland Blvd Caldwell, ID, 83605, USA}

\begin{abstract}
RR Lyrae stars are excellent tracers of stellar populations for old, metal-poor components in the Milky 
Way Galaxy and the Local Group.  Their luminosities have a metallicity-dependence, but determining 
spectroscopic $\rm [Fe/H]$ metallicities for RR Lyrae stars, especially at distances outside the solar 
neighbourhood, is challenging.  Using 40 RRLs with metallicities derived from both Fe(II) and Fe(I) abundances, we 
verify the calibration between the $\rm [Fe/H]$ of RR Lyrae stars from the Calcium triplet.  Our 
calibration is applied to all RR Lyrae stars with {\it Gaia} RVS spectra in {\it Gaia} DR3 as well as to 80 
stars in the inner Galaxy from the BRAVA-RR survey.  The co-added {\it Gaia} RVS RR Lyrae spectra 
provide RR Lyrae metallicities with an uncertainty of 0.25~dex, which is a factor of two improvement 
over the {\it Gaia} photometric RR Lyrae metallicities.  Within our Galactic bulge RR Lyrae star sample, we find a dominant fraction with low energies without a prominent rotating component.  Due to the large fraction of 
such stars, we interpret these stars as belonging to the $in-situ$ metal-poor Galactic bulge component, although we 
can not rule out that a fraction of these belong to an ancient accretion event such as Kraken/Heracles. 
\end{abstract}

\section{Introduction} 
RR Lyrae stars (RRLs) are some of the few objects for which distances can be reliably inferred.  
They were the first stars that provided a distance to the center of the Galaxy, a fundamental parameter of our Milky Way Galaxy \citep{blanco84}, are used to probe the most distant stars known in the Milky Way halo \citep[e.g.,][]{medina24, feng24} and their presence is used to find new low-luminosity globular clusters, substructures and streams \citep[e.g.,][]{sesar17, prudil21, cook22, butler24}.  RRLs are stars residing on the horizontal branch (HB), undergoing core helium burning.  
Because they reside on the part of the HB that intersects with the Cepheids instability strip, they pulsate 
(radially).  
Their well-known pulsation cycle, with periods that range from $\sim$0.2 to 1.0~days and with pulsation amplitudes of $\sim$1~mag in optical passbands, ensures they can be distinguished from other stars. 

Pulsation properties of stars on the main instability strip are linked to their physical parameters, such as their luminosity and metallic content.  
Period-luminosity relations of RRLs are therefore often used to determine distances to the star.  
That the absolute magnitude is a function of the $\rm [Fe/H]$ metallicity for the $V$-passband 
has been seen in stellar pulsation models \citep[e.g.,][]{christy66, marconi15}, 
stellar evolution tracks \citep{sweigart78, demarque00}, 
and also in observational studies \citep[e.g.,][]{liu90, cacciari92, carney92, fernley93, sandage93, mcnamara97, fernley98, muraveva18}.  In general, only $\rm [Fe/H]$ metallicity is taken into account when determining an RRL absolute magnitude ($M_V$) using optical passbands \citep[e.g.,][]{catelan04}.

In contrast to optical passbands, RRL absolute magnitudes in infrared (IR) passbands have 
been shown to be more affected by the pulsational period of the RRLs than their 
metallicity \citep[e.g.,][]{longmore86, catelan04,  dallora04, prudil24a}.  Further, both theoretical models 
and observational  studies show that the period-luminosity (PL) relations have a decreased scatter with 
increasing wavelength \citep[e.g.,][]{bono03, madore12, marconi15, vivas17}.  But even with the 
longer passbands, incorporating a metallicity term in the RRL PL relations improves the derived 
distance precision by at least 1-2 per cent \citep[e.g.,][]{neeley19, muhie21}.  

Photometric observations of RRLs allow periods to be determined with high accuracies, especially 
since most of the pulsation cycle of an RRL can typically be monitored over 12-18 hours.  
Unfortunately this is not the case for the $[\rm Fe/H]$ metallicity of an RRL, and typically the 
main source of uncertainty when using RRLs to obtain distances is the metallicity component 
in the period--luminosity--metallicity (PLZ) relation.  Exceptions to this are when it is reasonable to 
assume a single metallicity for the old stellar component, such as for cluster RRLs or for RRLs 
belonging to dwarf Spheroidal galaxies or ultra-faint dwarf galaxies, and metallicities from 
$e.g.,$ red giants' spectra can be adopted.  
The three main methods to determine RRL metallicities are from high-resolution spectroscopy, from low-resolution spectroscopy and/or from the shape of the RRLs photometric light curve. 

High-resolution spectroscopy (R$\gtrsim$25 000) provides a direct measurement of the metallicity of an RRL often 
with the highest accuracy and smallest $\rm [Fe/H]$ uncertainty than the other methods mentioned above.
Unfortunately, high-resolution spectroscopy is typcially feasible only for RRLs within a few kiloparsecs from the sun, as RRL absolute magnitudes ($M_V \sim$0.7) make spectroscopic observations of RRLs in the Galactic bulge and beyond both challenging and resource-intensive.  
There are $\sim$180 RRLs with metallicities determined from 
high-resolution spectroscopy \citep[e.g.,][]{dambis13, liu13, fabrizio19, gilligan21, crestani21a}.  
These abundances are on the \citet{for11, chadid17, sneden17, crestani21a} metallicity scale, abbreviated CFCS.  
In order to probe metallicities for RRLs in the outer halo of the MW, the Bulge, or in external Galaxies, lower-resolution spectroscopy or photometric metallicities, $\rm [Fe/H]_{phot}$, need to be employed.

Lower-resolution spectroscopy relies mainly on the $\Delta$S method to obtain a $\rm [Fe/H]$ 
metallicity \citep[e.g.,][]{preston59, clementini91}.  This method consists of comparing the spectral 
type or equivalent widths (EWs) of the RRL using one family of lines (e.g., the hydrogen lines, 
H$_\beta$, H$_\gamma$, H$_\delta$), to the 
spectral type  or equivalent width of the RRL using a different line (e.g., the Ca {\tt II} K line).  
These Hydrogen and Calcium lines are strong and dominate the spectrum, making them straight-forward 
to identify and use as tracers of $\rm [Fe/H]$ metallicity.  This $\Delta$S -- the difference between the 
spectral type or EWs from different spectral lines -- is directly correlated to the $\rm [Fe/H]$ metallicity of an 
RRL \citep[e.g.,][]{butler75, butler82, walker91, layden94}.  
\citet{wallerstein12} showed that the $\Delta$S method can be used with only the equivalent 
width of the Ca {\tt II} line at 8498 \AA~to derive a RRL $\rm [Fe/H]$ metallicity.  This is especially useful 
if the wavelength regime is insufficient to cover both Ca {\tt II} lines as well as H$_\beta$, H$_\gamma$, 
H$_\delta$ lines. 

The use of photometric metallicities for RRLs relies on an empirical relation between mathematical 
coefficients that measure the RRL 
light-curve shape, the period and the $\rm [Fe/H]$ metallicity.  A RRL stars light-curve shape is frequently 
quantified by a Fourier decomposition to the RRLs periodic photometric signal \citep{jurcsik96}, but 
other mathematical coefficients that describe the shape of the RRLs lighturve have also been used \citep[e.g., using a Principal-Component Analysis, PCA,][]{kanbur04}.  A recent study utilizing the \citet{crestani21a} high-resolution 
metallicities combined with thousands of lower-resolution metallicities from LAMOST DR6 and 
SDSS-SEGUE, show that the intrinsic scatter in the photometric metallicities is 0.41 dex in the $V$-band 
and 0.50 dex in the infrared \citep{mullen21}.  

Here we focus on using lower-resolution spectroscopy covering the Ca {\tt II} lines to probe the 
$\rm [Fe/H]$ metallicities of RRLs, $\rm [Fe/H]_{CaT}$.  The Calcium lines were first used as metallicity indicators 
because the Ca {\tt II} IR triplet is red enough to reduce the effect of interstellar extinction and hence 
to allow heavily reddened clusters in the MW to be spectroscopically analyzed \citep[$e.g.,$][]{armandroff88}.  
Since then, the use of Ca {\tt II} lines to measure metallicity has especially shown to be powerful for 
metal-poor stars \citep[$e.g.,$][]{starkenburg10, carrera13, matijevic17}, as for low metallicity 
stars ($\rm [Fe/H] < -$2.0), many metallic lines become almost indistinguishable from the background 
noise, but the Ca {\tt II} lines are still prominent.  

The calibration of the EW of the 8498\AA~Ca {\tt II} line in RRLs to $\rm [Fe/H]$ metallicity was first 
presented by \citet{wallerstein12}, with the goal to use this calibration to derive metallicities of RRLs that 
would be observed by {\it Gaia}.  Only the 8498\AA~Ca {\tt II} line was used as the others were often blended by 
neighboring Hydrogen Paschen lines.  The calibrating sample is based on R$\sim$35 000 spectra from the 
Apache Point Observatory (APO) echelle spectrograph of $\sim$30 RRLs.  This relation has been applied to 
935 RRLs in the Galactic bulge observed in the BRAVA-RR survey \citep{kunder16, kunder20} to obtain the 
first spectroscopic metallicity distribution function for a large sample of RRLs in the Galactic bulge \citep{savino20}.  
The global metallicity distribution function was shown to be good to 0.35~dex, but individual RRL 
$\rm [Fe/H]_{CaT}$ metallicities could have much larger uncertainties, hampered mainly by the low 
signal-to-noise (S/N) of the BRAVA-RR spectra.  

Ca {\tt II} abundances have not been applied to the {\it Gaia} RRLs yet, but this has tremendous potential 
considering the plethora of {\it Gaia} spectra that are, or will be, available with ongoing {\it Gaia} data 
releases.  There also exists the possibility to obtain $\rm [Fe/H]_{CaT}$ for individual BRAVA-RR 
stars with uncertainties that allow for a chemodynamical mapping of the stellar population in the inner Galaxy.  
In \S2 we expand on the \citet{wallerstein12} EW-$\rm [Fe/H]$ calibration and verify its accuracy.  
$\rm [Fe/H]_{CaT}$ abundances for 175 {\it Gaia} RRLs are obtained and discussed in \S3, and an analysis of 
how $\alpha$ affects the {\it Gaia} photometric metallicities, $\rm [Fe/H]_{phot,DR3}$ of a star is carried out.  
$\rm [Fe/H]_{CaT}$ metallicities of 80 Galactic bulge RRLs are derived in \S4, and used to probe the signature of the 
earliest known Milky Way accretion event, Heracles.  Our conclusions are in \S5.

\section{Calibration of the Calcium Triplet} \label{sec:calibration}
Here we carry out an independent study from that by \citet{wallerstein12} to probe the dependence of the EW of the 8498\AA~Ca {\tt II} line in RRLs to their $\rm [Fe/H]$ metallicity.  
Fundamental-mode RRLs (RRab), first-overtone (RRc), and double-mode pulsators (RRd) with spectroscopic $\rm [Fe/H]$ abundances and photometric (pulsation properties) information is available in \citet{crestani21a}, and this sample was used as calibrators of the Ca \texttt{II}-[Fe/H] relation.  
The ESO archive\footnote{\url{https://archive.eso.org/scienceportal/home}} was used to access the RRL spectra from the UVES ($R \approx 40000$) and XSHOOTER ($R \approx 10000$) instruments. Additionally, spectra from the RAVE survey \citep[$R \approx 7500$,][]{steinmetz20} was obtained for a handful of RR Lyrae stars with $\rm [Fe/H]$ in the \citet{crestani21a} compilation. In total, $68$ spectra for $40$ single-mode pulsators were collected and for each spectrum, we obtained its signal-to-noise ratio ($\rm S/N$), resolution, exposure, and observation time. 
The $\rm S/N$ of the RRL spectra varied between $28$ to $222$, with a median centered at $69$.  The spectra collected are listed in Table~\ref{tab:CaTcalib} where the columns list each RRLs {\it Gaia} ID (1), $\rm [Fe/H]$ from \citet{crestani21a} (2), period (3), time of maximum brightness (4), time the spectra was taken (5), type of single-mode RRL (6), spectra name from archive (7), S/N (8), resolution (9), instrument (10), measured EW and uncertainty about the measurement (11).  

To estimate equivalent widths (EW) for Ca \texttt{II} at $8498$\,\AA~from each spectrum, a {\tt Python} routine was developed that allows the user to select continuum and line regions based on the properties 
of individual spectra in order to measure the equivalent widths (EW) for Ca \texttt{II} at $8498$\,\AA~from each spectrum (see example fit for {\it Gaia} RVS spectra in Figure~\ref{fig:CaTcode}). 
The routine works with a pre-synthesized spectrum using the \texttt{iSpec} module \citep{blancocuaresma14, blancocuaresma19} for typical RR~Lyrae stars ($\rm [Fe/H]=-1.5$\,dex, $T_{\rm eff}=6500$\,K, log~$g=2$\,dex) that serves as a guide to locate continuum regions in the spectrum. 
As in \citet{wallerstein12}, only the EWs for the Ca {\tt II} 8498\,\AA~lines were measured, as this line is the furthest from the hydrogen line of the Paschen series.

\begin{figure}
\centering
\mbox{\subfigure{\includegraphics[height=6.2cm]{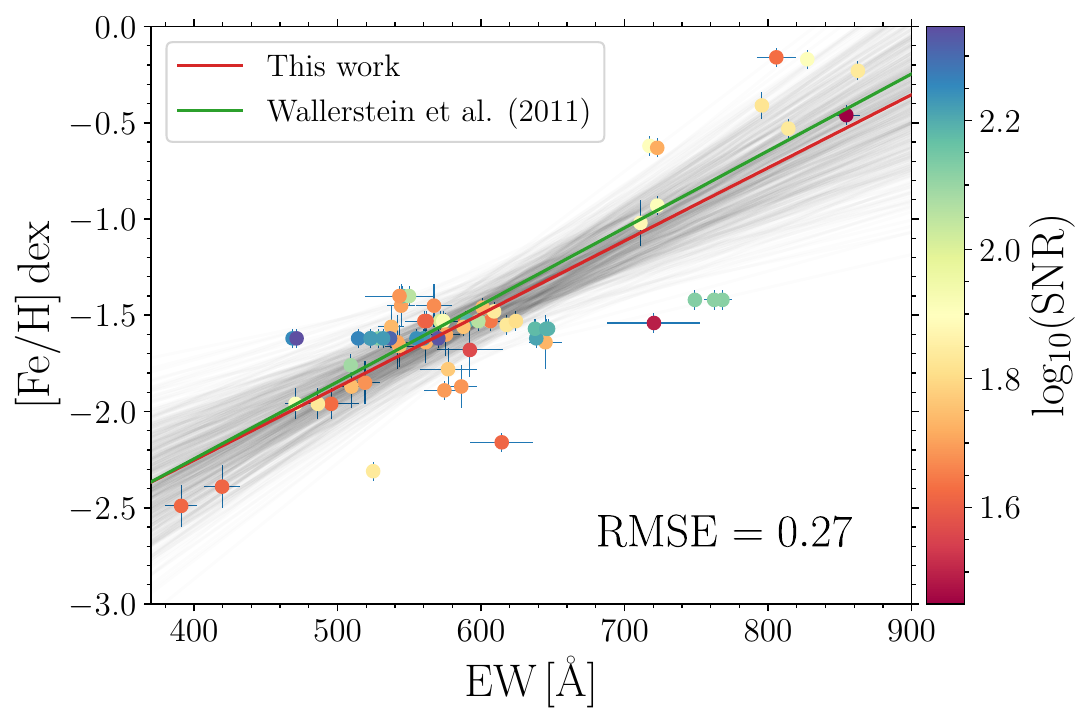}}}
\caption{The calibrated linear relation (red line) between spectroscopic $\rm [Fe/H]$ and EW of the Ca \texttt{II} at $8498$\,\AA~line. The color-coding illustrates the S/N of the individual measurements.  The individual measurements with all their available measurements are shown, so some individual RRLs are represented here repeatedly.
The green line represents the calibration relation from \citet{wallerstein12}.}
\label{fig:CalibrationEWFEH}
\end{figure}

Figure~\ref{fig:CalibrationEWFEH} shows the correlation between our measured EWs for Ca \texttt{II} at $8498$\,\AA~and the $\rm [Fe/H]$ metallicities from \citet{crestani21a}.  
Both our measured EWs and the $\rm [Fe/H]$ abundances were derived with their associated uncertainties, so our fitting procedure considers both sources of error when determining the best relation between the two.  
Because both of these quantities were derived independently, we assume that the correlation between them is negligible. The utilized covariance matrix $\Xi^{i}$ for a given star ($i$) has the following form:
\begin{equation}
\Xi^{i} = \begin{bmatrix}
\sigma_{\text{EW}}^{i} & 0 \\ 0 & \sigma_{\text{[Fe/H]}}^{i} \\ \end{bmatrix} \\.
\end{equation}

To obtain the best fit, the procedure by \citet[][see their Sec.~7]{hogg10} was followed, an approach that is fairly similar to the orthogonal least square method. In this method, the slope of the linear relation, $a$ is described by a unit vector $\mathbf{v}$ perpendicular to the linear fit:

\begin{equation}
\mathbf{v} = \frac{1}{\sqrt{1+a^{2}}} \begin{bmatrix}
-a \\ 1
\end{bmatrix}  \\.
\end{equation}

The slope is defined by an angle $\theta$ via $a = \text{tan}(\theta)$. Each data point $i$ (EW$^{i}$ and [Fe/H]$^{i}$) can be represented by a vector $\mathbf{Z}_{i} = (\text{EW}^{i}, \text{[Fe/H]}^{i})$. The orthogonal displacement $\delta_{i}$ for each point is then:

\begin{equation}
\delta_{i} = \mathbf{v}^{\rm T} \times \mathbf{Z}_{i} - b\cdot \text{cos}(\theta) \\,
\end{equation}

where the parameter $b$ represents the intercept of the linear fit. In addition to the slope and intercept, we included $\varepsilon$, representing the intrinsic scatter in the EW vs. [Fe/H] relation. The orthogonal variance, $\Sigma_{i}^{2}$ is then described by:

\begin{equation}
\Sigma_{i}^{2} = \mathbf{v}^{\rm T} \times \Xi_{i} \times \mathbf{v} + \exp{\varepsilon}\\.
\end{equation}

The aforementioned parameters then form the following relation for the log-likelihood function $\text{ln} (p)$:

\begin{equation}
\text{ln} (p) = K \sum_{i}^{N} \frac{\delta_{i}^{2}}{2\Sigma_{i}^{2}} \\.
\end{equation}

\noindent 
where $K$ is a constant.  The log-likelihood function can be optimized for the linear model parameters $a$ and $b$, yielding the best values for $a$, $b$, and $\varepsilon$. In the maximization of the log-likelihood, we utilized the \texttt{emcee} module \citep{foremanmackey13}, and we ran the Markov Chain Monte Carlo simulation with $200$ walkers for $10000$ samples. For deriving the best-fit values, we thinned the sample by $\tau=100$ and marked the initial $5000$ samples as burn-in.

In this way, the following relation and covariance matrix was obtained:
\begin{equation} 
\label{eq:CalibrRel}
\text{[Fe/H]}_{\text{Ca} \texttt{II}} = 0.0038 \cdot \text{EW} - 3.77 \\
\hspace{4.0cm}
\text{Cov} = 
\begin{bmatrix} 
$0.0000014$ & $-0.0008481$ \\
$-0.0008481$ & $0.5160122$ \\
\end{bmatrix}
\hspace{0.3cm} \varepsilon = 0.04 \hspace{0.1cm} .
\end{equation}

This relation has a root-mean-square-error (RMSE) of 0.27~dex, and is remarkably similar to that presented in \citet{wallerstein12}.  
Although our relation was derived using RRab, RRc and RRd pulsators, there are more RRab-type RRLs in our calibration sample than RRc or RRd stars, so the calibration is most appropriate for RRab stars.  We examined how the quality of the spectra, as represented by the signal-to-noise ratio (S/N), influenced our calibration by applying several S/N cuts. The resulting relation in Eq.~\ref{eq:CalibrRel} does not exhibit significant changes within the one $\sigma$ errors listed in its covariance matrix. Overall, the quality of the selected spectra is high and has a negligible influence on the final calibration.  Because all of our calibration RRLs are from the \citet{crestani21a} sample, this calibration is on the CFCS metallicity scale.

\section{{\it Gaia} RR~Lyrae~star Metallicities}
Individual photometric metallicities for {\it Gaia} RRLs presented in {\it Gaia} DR3 come from the Specific Object Study (SOS) pipeline.  The SOS module that provides the community with a photometric $\rm [Fe/H]$ metallicites uses the RRLs pulsation period, $P$, and 
the $\phi_{31}$\footnote{$\phi_{31} = \phi_{3} - 3 \phi_{1}$, where $\phi_k$ are Fourier phases from a sine-series sum:
m(t) = $A_0 + \sum_{k=1}^N A_k\sin (k2\pi t/P + \phi_k)$ } parameters of the 
$G$ light-curve Fourier decomposition with the equation derived in \citet{nemec13}:
\begin{equation}
\label{eqn:photFe}
\rm [Fe/H]_{phot,DR3} = -5.241 - 5.394P + 1.345~\phi_{31} \\
\end{equation}
In this way, metallicities are available for 133 557 RRLs in the {\it Gaia} data archive.  
\citet{clementini23} validate these photometric metallicities by comparing with the 
high-resolution metallicities of \citet{crestani21a}.  
They find a mean photometric metallicity error of 0.46~dex.

The {\it Gaia} all-sky RRL catalogs have also generated interest  in developing photometric metallicities in the $G$-band \citep{iorio21, li23, dekany22, jurcsik23, clementini23}.
None of the photometric metallicity relations put forward to date, however, use the large, homogeneous spectroscopic survey of field RRLs from \citet{crestani21a}, which covers more than three dex in iron abundance, as a calibrating sample.  We believe this sample has a great potential to anchor a photometric metallicity calibration in both the metal-poor and metal-rich end of the RRLs, and that the homogeneity of the analysis of this sample can reduce effects of intrinsic scatter in the calibrating sample.  
Therefore, a calibration of the photometric metallicity using the {\it Gaia} $G$ passband is carried out for both RRab and RRc stars using the \citet{crestani21a} stars as calibrators.  These results can be found in Appendix~\ref{sec:appendixA} and are summarized in Table~\ref{tab:review_phot_feh}.

Table~\ref{tab:review_phot_feh} also summarizes additional relations put forward for photometric $\rm [Fe/H]$ metallicities from previous studies.  
\citet{iorio21} use 84 RRLs from the spectroscopic sample of \citet{layden94} to calibrate a photometric 
$\rm [Fe/H]$ for RRLs with an instrinsic scatter of 0.31~dex for the RRab stars and 0.16~dex for RRc stars. 
\citet{li23} use a catalog of $\sim$5290 RRLs from \citet{liu20} to calibrate $\rm [Fe/H]_{phot}$ 
using the {\it Gaia} photometric properties of their observed RRLs.   They find a scatter of 0.24~dex 
for the RRab relation and 0.19~dex for their RRc relation.  
\citet{dekany22} use deep learning and the Gaia $G$-band light curves that have a near-infrared $K_s$-band 
light curve counterpart to train a neural network for a $\rm [Fe/H]_{phot}$ estimation.  The results from this 
process was then applied to other passbands, such as the Gaia $G$-band, and was found to have a scatter 
comparable to the \citet{crestani21a} high-resolution uncertainties.
\citet{jurcsik23} use the {\it Gaia G} band RRL light curves that overlap with a globular cluster (GC) RRL star to assemble a sample 
of 526 high-quality RRLs belonging to 70 GCs.  The high-resolution, well-studied spectroscopic $\rm [Fe/H]$ 
metallicities of the GCs are used as calibration $\rm [Fe/H]_{phot}$ metallicities.  They find that their relation 
does not work for Oosterhoff II-type RRLs -- RRLs with longer periods for their pulsation amplitudes -- and they do not believe there is a photometric $\rm [Fe/H]$ relation that is accurate for Oosterhoff II RRLs.  Note that local 
RRLs in general are Oosterhoff I-type RRLs; it is the more metal-poor RRLs that tend to be Oosterhoff II-type stars \citep[e.g.,][]{prudil19, prudil20}.

\begin{deluxetable*}{llcccc}
\tablenum{2}
\tabletypesize{\scriptsize}
\tablecaption{{\it Gaia} photometric $\rm [Fe/H]$ metallicity relations
\label{tab:review_phot_feh}}
\tablehead{
\colhead{Authors} & \colhead{Photometric $\rm [Fe/H]$ Relation} & \colhead{$\sigma_{\rm published}$ } & \colhead{$\sigma_{\rm CaT, all~S/N}$ } & \colhead{$\sigma_{\rm CaT, S/N > 35}$} & \colhead{Correlation in [$\alpha$/Fe] }  
} 
\decimalcolnumbers
\startdata
\citet{clementini23} & RRab: $-$5.241 $-$5.394~$P$ + 1.345~$\phi_{31}$ & 0.46 & 0.52 & 0.44 & $-$0.74$\pm$0.32 \\ 
 & RRc: \ \ $-$5.241 $-$5.394~$P$ + 1.345~$\phi_{31}$ & 0.46 & 0.44 & 0.37 &  \\ 
  \hline
\citet{iorio21} & RRab: $-$1.68 $-$5.08~($P$-0.6) + 0.68~($\phi_{31}$-2.0) & 0.31 & 0.34 & 0.22 & $-$1.43$\pm$0.18 \\ 
  & RRc: \ \ $-$1.26 $-$9.39~($P$-0.3) + 0.29~($\phi_{31}$-3.5) & 0.16 & 0.37 & 0.34 &  \\
  \hline 
\citet{li23} & RRab: $-$1.888 $-$5.772~($P$-0.6) + 1.09~($\phi_{31}$-2.0) & 0.24 & 0.36 & 0.22 & $-$0.95$\pm$0.22  \\
  & \ \ \ \ \ \ \ \ \ \ + 1.065~($R21$ - 0.45) &  &  &  &  \\ 
  & RRc: \ \ $-$1.737 $-$9.968~($P$-0.3) + -5.041~($R21$-0.2) & 0.19 & 0.33 & 0.31  &   \\
  \hline 
\citet{dekany22} & RRab: Neural Network & 0.21& 0.42 & 0.26 & $-$0.81$\pm$0.17 \\ 
   \hline 
\citet{jurcsik23} & RRab: $-$3.504+3.14 $-$5.716~$P$ + 1.019~$\phi_{31}$  & 0.21 & 0.43 & 0.20 & $-$1.04$\pm$0.21 \\ 
   \hline 
This Work & RRab: $-$5.5061~$P$ + 0.8143~$\phi_{31}$ - 0.0813  & 0.25 & 0.35 & 0.21 & $-$0.95$\pm$0.19 \\ 
  & RRc: \ $-$13.2023~$P$ + 0.5138~$\phi_{31}$ + 0.7338 & 0.19 & 0.44 & 0.40  & \\
   \hline %
\enddata
\end{deluxetable*}

\subsection{RRL Photometric Metallicities compared to Calcium Triplet Metallicities}

The brighter stars observed by the {\it Gaia} satellite ($G \sim$16) have medium-resolution spectra 
delivered by the Radial Velocity Spectrometer (RVS, [845 -- 872]~nm, 
$\lambda$/$\Delta \lambda \sim$1500).  In the current {\it Gaia} data release, DR3, there are 213 {\it Gaia} 
RRLs with RVS spectra available for download ({\tt gaia\_source.has\_rvs=true}).  
However, 36 of those have no clean RV epochs ({\tt num\_clean\_epochs\_rv = `null'}) and correspondingly have 
no average radial velocity; these are not used in our analysis.  Therefore, our final 
$Gaia$ RRL RVS sample consists of 177 stars.  Only co-added 
spectra is available for download, and therefore the {\it Gaia} spectra contain contributions from all the observed phases of the RRL in their Ca lines.  
Whereas a discussion on how the EW of the CaT changes as a function of phase can not be carried out using the {\it Gaia} RVS spectra, we are able to use single-epoch spectra from the BRAVA-RR survey to examine the effect of pulsation phase on $\rm [Fe/H]_{CaT}$ and this is discussed in \S\ref{sec:bravarr}.

The number of spectra stacked to make the mean {\it Gaia} RVS spectra ranges from seven to fifty-four co-added exposures and the S/N of the spectra ranges from 20 to more than 80 for the brightest RRLs.  
The top right panel in Figure~\ref{gaia_catfeh} shows an example of the co-added RVS spectra of a sample of {\it Gaia} RRLs with different S/N ratios.

\begin{figure*}
\centering
\mbox{\subfigure{\includegraphics[height=10.0cm]{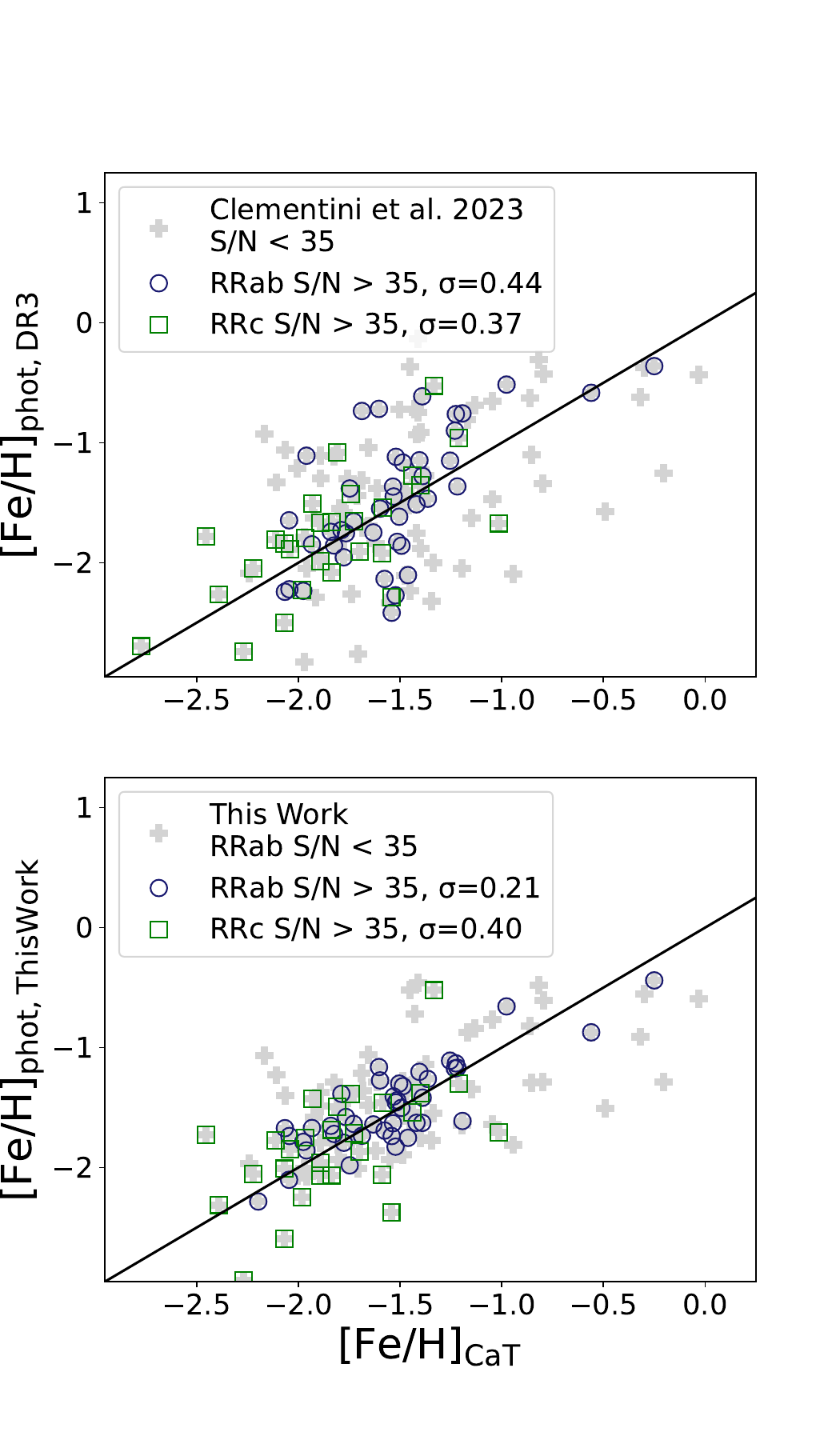}}}
{\subfigure{\includegraphics[height=10.0cm]{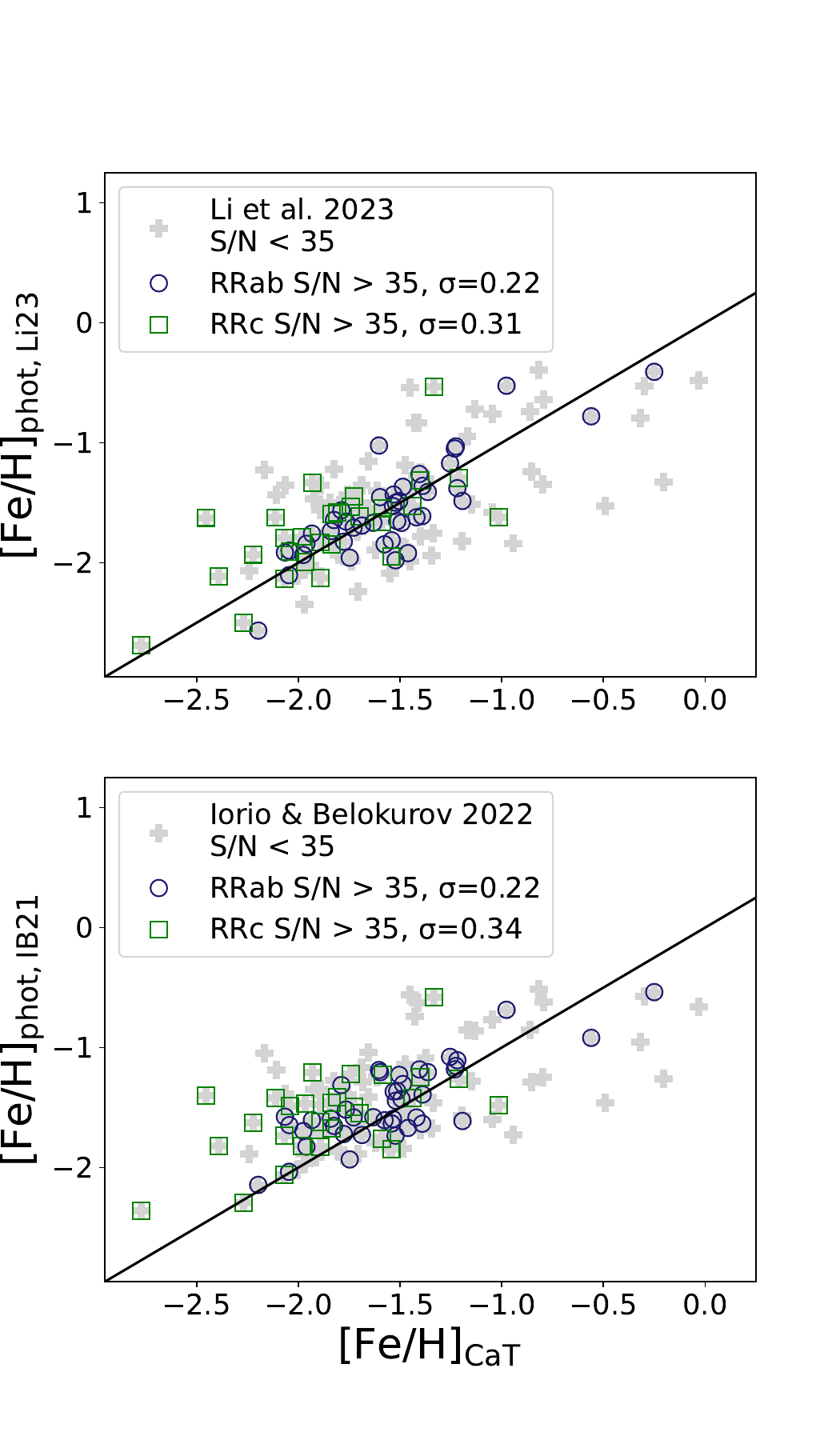}}}
{\subfigure{\includegraphics[height=10.0cm]{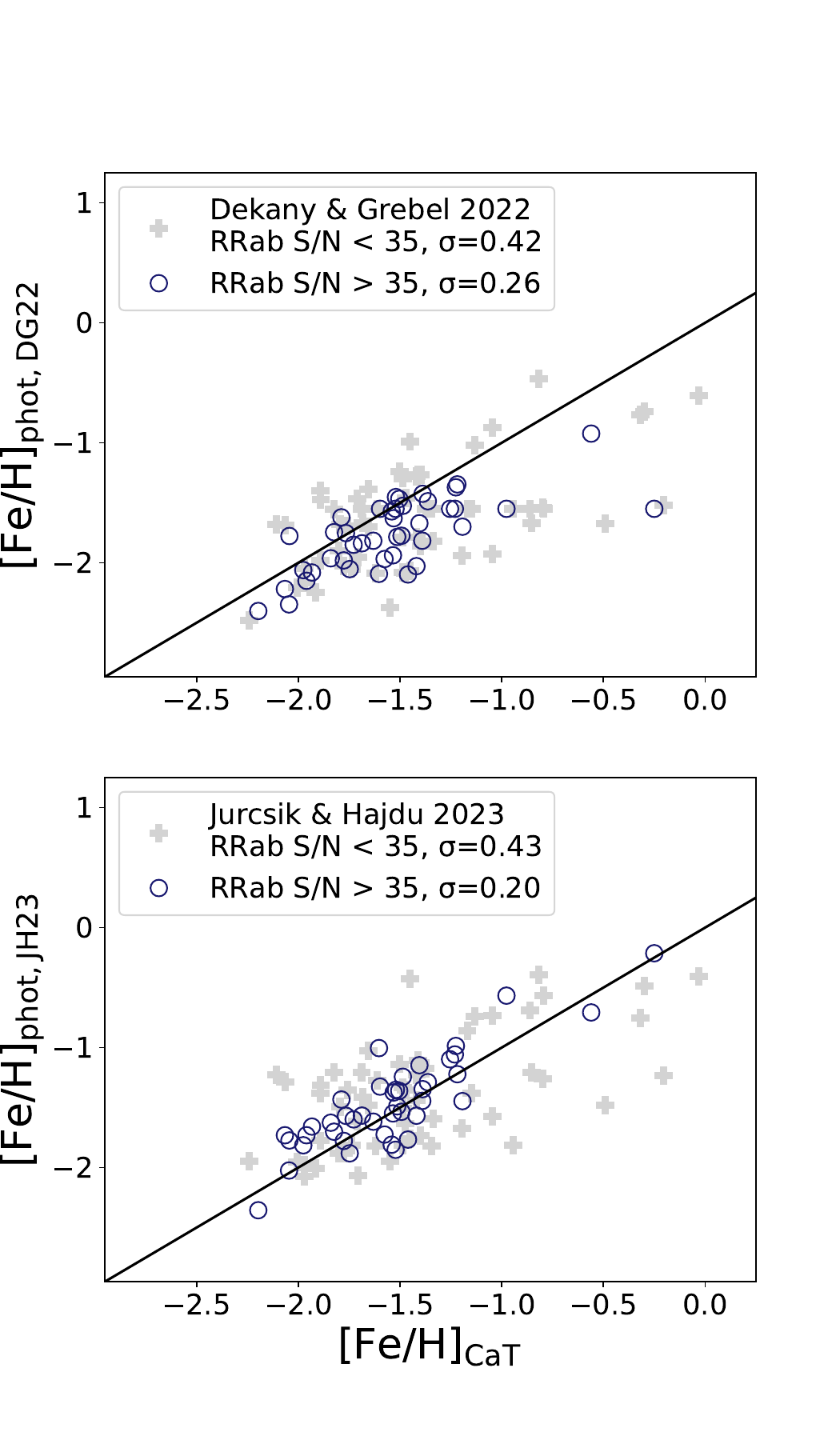}}}
\caption{
The difference between the $\rm [Fe/H]$ metallicity determined from the 8498\AA~Ca {\tt II} line and the 
$\rm [Fe/H]$ metallicity determined from different photometric $G$-band lightcurve-metallicity relationships 
in the literature.
The dispersion for the RRLs with RVS spectra with S/N $>$ 35 and S/N $<$ 35 is indicated in each panel.  
Our CaT metallicities have dispersions comparable to the precision of the photometric $\rm [Fe/H]$ metallicities 
for RRLs with S/N $>$ 35. The bottom left panel is photometric metallicity from this study where the calibration is described in the Appendix~\ref{sec:appendixA}.
}
\label{fehphotcomps}
\end{figure*}

The dispersion between our $\rm [Fe/H]_{CaT}$ metallicities and the published photometric metallicities is reported in Table~2, and Figure~\ref{fehphotcomps} shows the difference between our $\rm [Fe/H]_{CaT}$ metallicities and the photometric metallicities relations.  
A total of 137 RRLs can be compared, as 38 RRLs with RVS spectra do not have light curves with sufficient {\it Gaia} photometric measurements for a $\phi_{31}$ value (and hence do not have a photometric $\rm [Fe/H]$ metallicity) and 2 RRLs did not have spectra that enabled a robust EW measurement.  
The \citet{dekany22} and \citet{jurcsik23} photometric metallicities are only derived for RRab stars, and so only the 100 RRab stars are shown.  
The photometric metallicity relation presented here (Appendix~\ref{sec:appendixA}), as well as those from \citet{iorio21}, \citet{li23} and \citet{clementini23} are given for both RRab and RRc stars, and so 100 RRab as well as 37 RRc stars are displayed.  

Here it is evident that the $\rm [Fe/H]_{CaT}$ is a function of S/N as well as pulsation mode ($e.g.,$ RRab or RRc).  $\rm [Fe/H]_{CaT}$ metallicities agree better with photometric metallicities for those stars with higher S/N spectra, and also, for RRab stars rather than for RRc stars.  The dispersion between $\rm [Fe/H]_{phot}$ and $\rm [Fe/H]_{CaT}$ is given both for the RRab and RRc stars with RVS spectra with S/N $>$ 35.  

The dispersion between the $\rm [Fe/H]_{phot,DR3}$ and $\rm [Fe/H]_{CaT}$ is 
$\sigma =$0.41 for the 70 stars with S/N $>$ 35, which is comparable to the mean photometric metallicity 
error.  In contrast, the dispersion between the photometric and CaT $\rm [Fe/H]$ metallicities is 
$\sigma =$0.52 for the 65 stars with S/N $<$ 35.  
The scatter between $\rm [Fe/H]_{CaT}$ and the independent $\rm [Fe/H]_{phot}$ values show an improved agreement as compared to $\rm [Fe/H]_{phot,GaiaDR3}$, indicating that the uncertainty in our $\rm [Fe/H]_{CaT}$ metallicities is at least comparable to $\rm [Fe/H]_{phot}$ for spectra with S/N $>$ 35.  
The uncertainties in $\rm [Fe/H]_{CaT}$ are further assessed using comparisons with high-resolution spectra below.

\subsection{{\it Gaia} Metallicities compared to High-Resolution Metallicities}

\begin{figure*}
\centering
\mbox{\subfigure{\includegraphics[height=7.1cm]{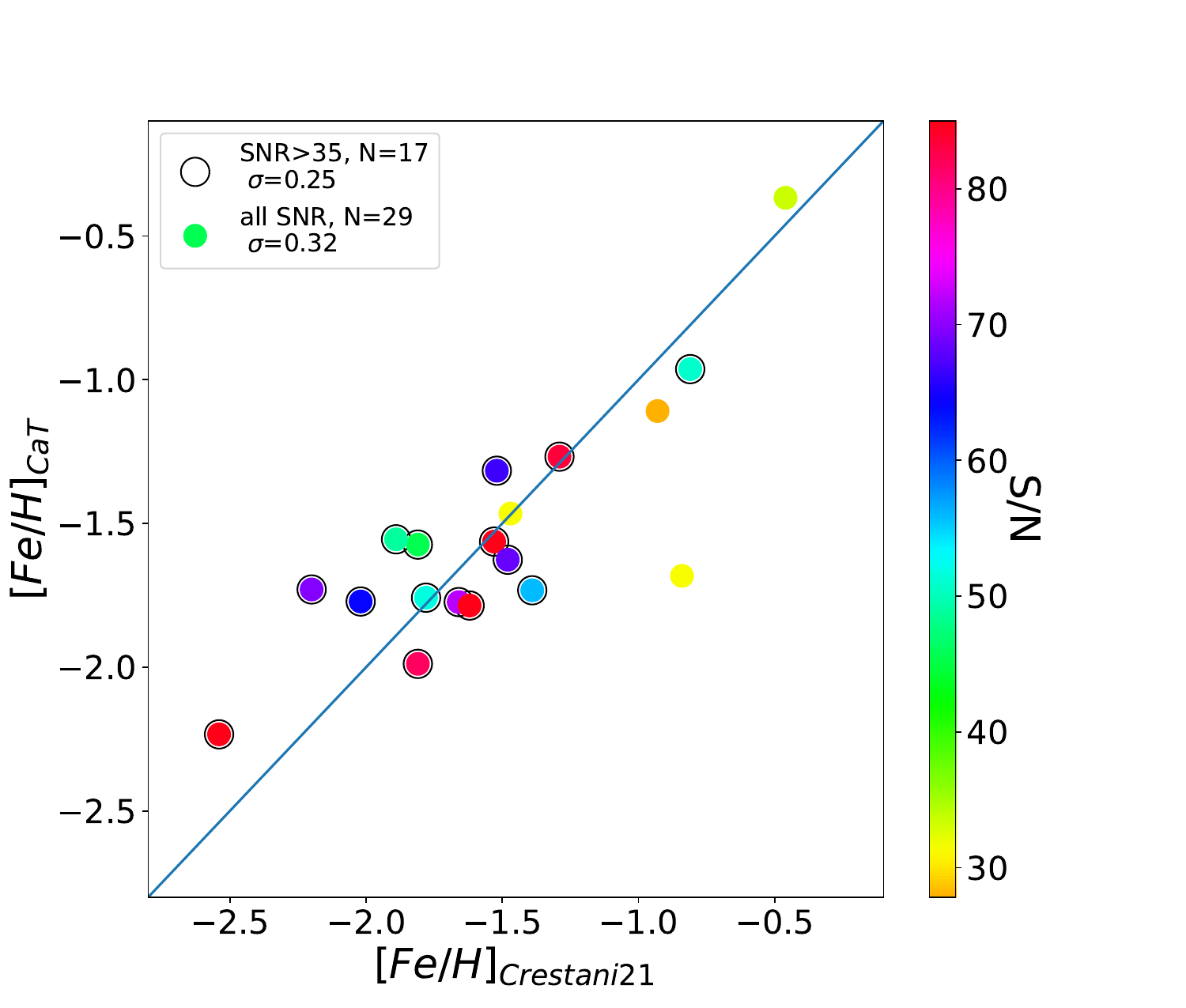}}}
{\subfigure{\includegraphics[height=7.1cm]{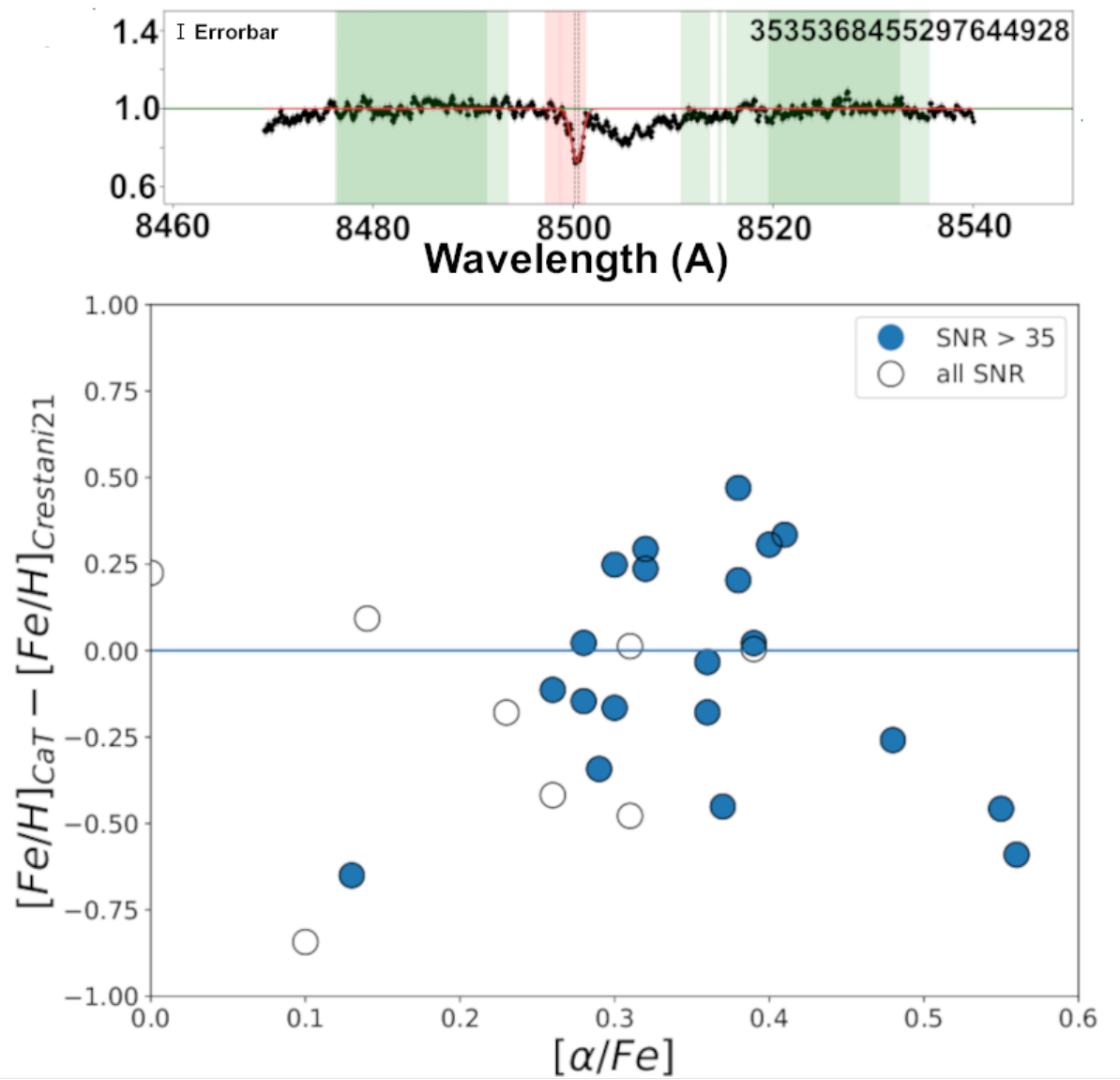}}}
\caption{
{\it Left:} The high-resolution $\rm [Fe/H]$ metallicities of \citet{crestani21a}, $\rm [Fe/H]_{Crestani21}$, 
compared to $\rm [Fe/H]_{CaT}$ of the same stars. 
{\it Bottom Right:} The $\rm [\alpha/Fe]$ abundance as a function of the difference between $\rm [Fe/H]_{CaT}$ and $\rm [Fe/H]_{Crestani21}$.  The $\rm [Fe/H]_{CaT}$ calibration does not appear signicantly affected by $\alpha$, but 
the sample size is small.
{\it Top Right:} 
The {\it Gaia} co-added RVS spectra of RRLs with S/N $\sim$ 35.  The blue and red continuum of the spectra are manually selected (green regions), as is the position of the CaT line at 8498~\AA, for which the EW is then measured.
}
\label{gaia_catfeh}
\end{figure*}

Twenty-two of our $Gaia$ RRL RVS sample were observed by \citet{crestani21a} 
in their high-resolution spectroscopy sample.  
The dispersion between 
the \citet{crestani21a} metallicities and those obtained from the CaT is $\sigma=$0.30 for the full sample, 
and $\sigma=$0.25 for the RVS spectra with S/N $>$ 35, as seen in Figure~\ref{gaia_catfeh}.
Because the \citet{crestani21a} compilation also include Ca, Ti and Mg elemental abundances, how 
alpha-abundances 
affect $\rm [Fe/H]_{CaT}$ can be investigated.  The right panel in Figure~\ref{gaia_catfeh} 
does not show indication that $\rm [\alpha/Fe]$ affects the derived $\rm [Fe/H]_{CaT}$, but this is still a small sample.  
It is further possible to use the \citet{crestani21a} alpha-abundances to investigate how $\alpha$ affects 
the {\it Gaia} photometric metallicities.  In this case the sample size will be larger, as $\rm [Fe/H]_{phot}$ is 
independent on the available RVS spectra.  

\begin{figure}
\centering
\mbox{\subfigure{\includegraphics[height=6.7cm]{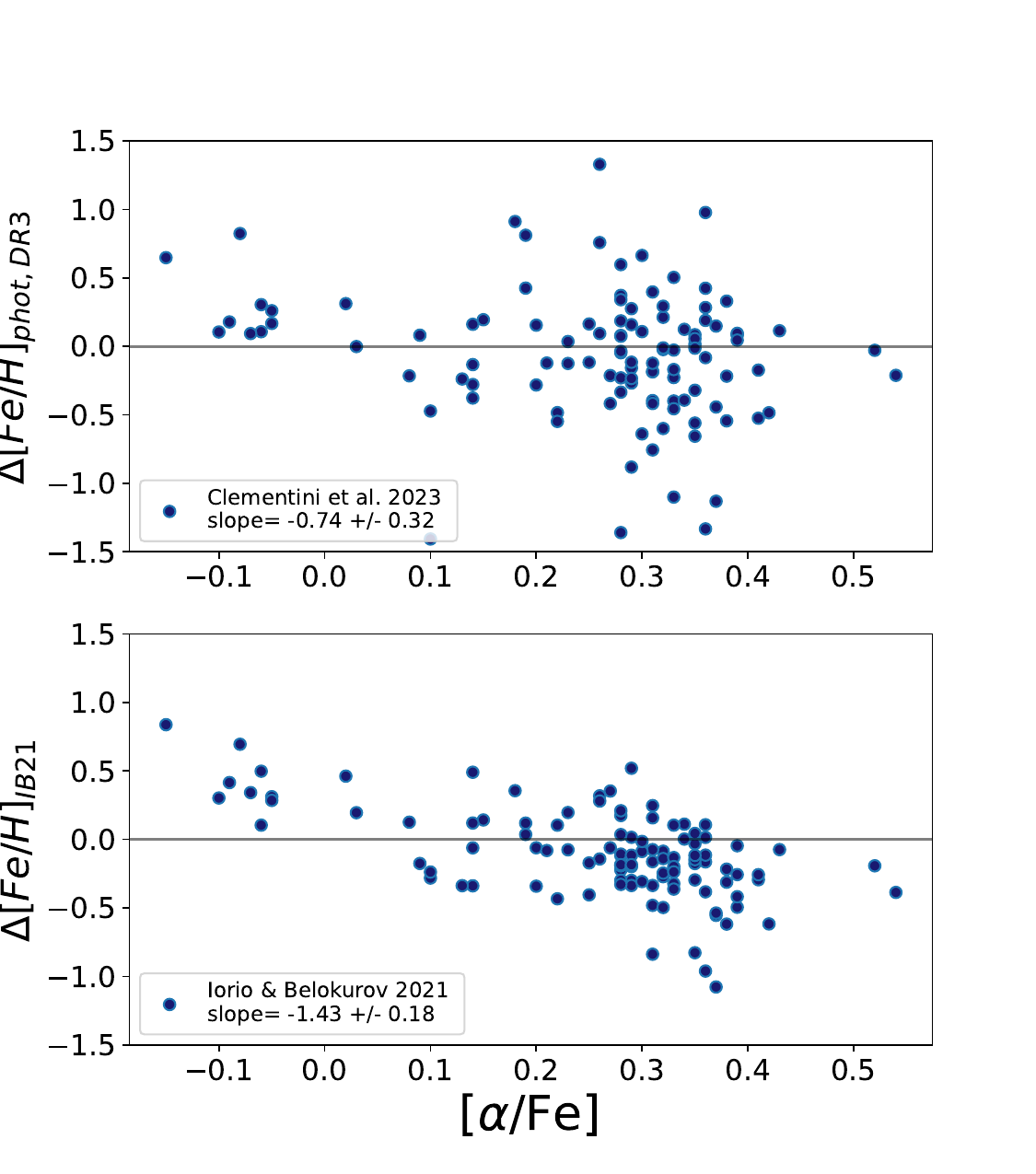}}}
{\subfigure{\includegraphics[height=6.7cm]{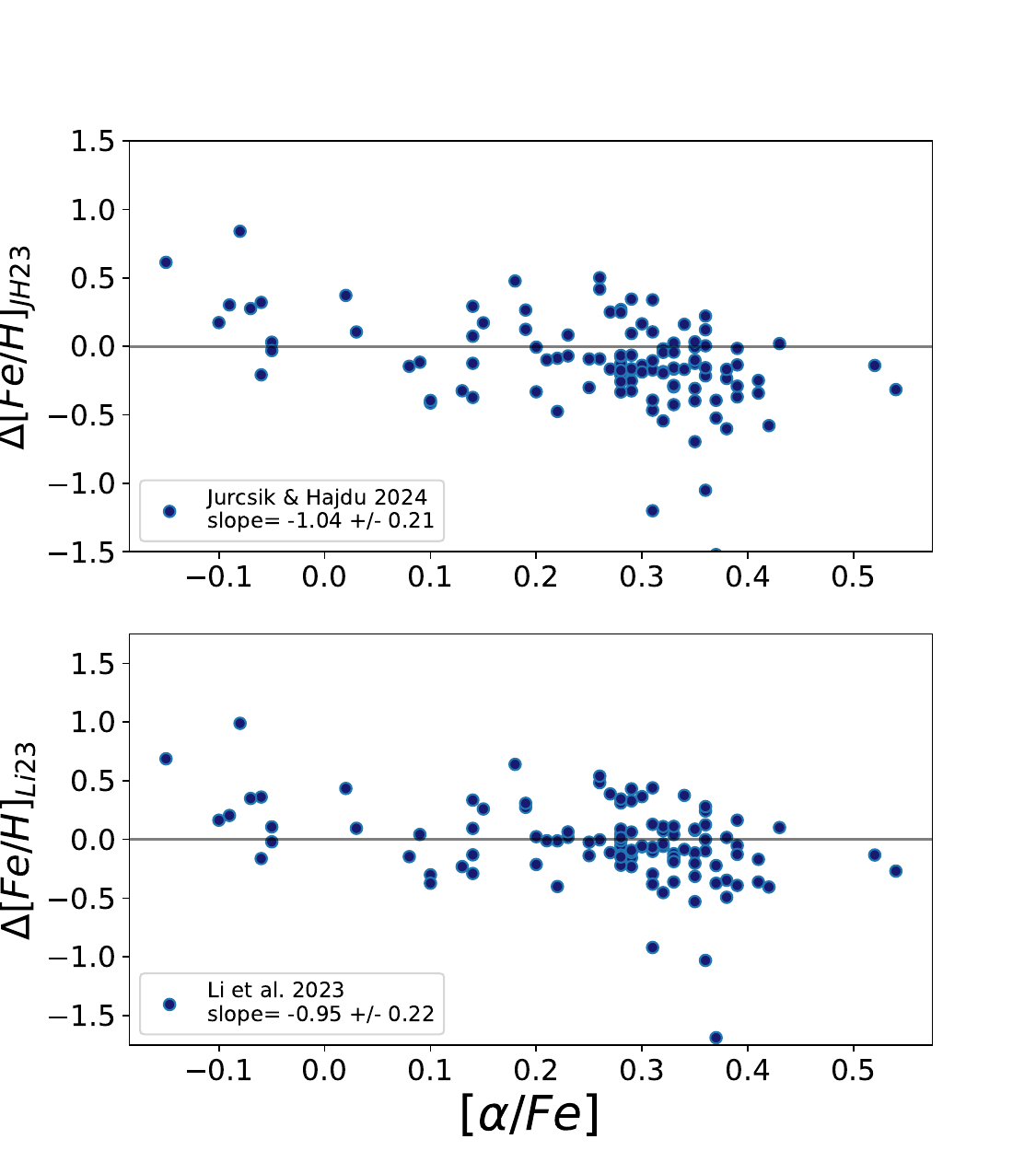}}}
{\subfigure{\includegraphics[height=6.7cm]{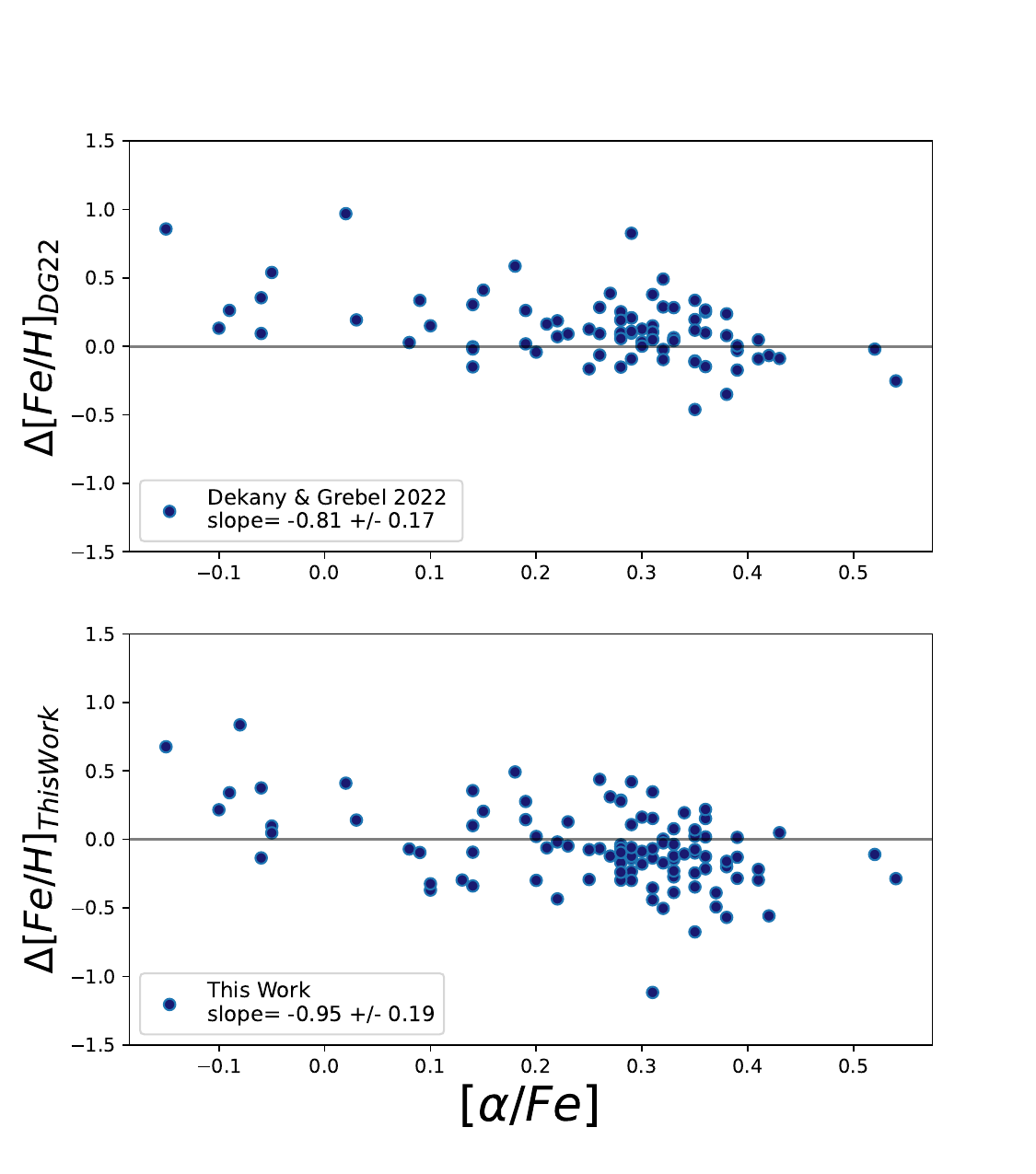}}}
\caption{
The $\rm [\alpha/Fe]$ abundance as a function of the difference between $\rm [Fe/H]_{CaT}$ and $\rm [Fe/H]_{phot}$, 
where the different $\rm [Fe/H]_{phot}$ relations of \citet{clementini23}, \citet{iorio21}, \citet{li23}, \citet{dekany22}, \citet{jurcsik23} and the study (Appendix~\ref{sec:appendixA}) are compared.  In all cases, $\rm [Fe/H]_{phot}$ is a function of the $\alpha$-enrichment of the RRL, with $\rm \Delta[Fe/H] >$ 0.5 indicating a star with a low $\rm [\alpha/Fe]$ abundance.
}
\label{alpha5}
\end{figure}

Figure~\ref{alpha5} shows how $\rm [Fe/H]_{phot}$ from different relations correlate to $\rm [\alpha/Fe]$, where $\rm \Delta [Fe/H]_{phot,DR3}$ corresponds to $\rm [Fe/H]_{CaT} - [Fe/H]_{phot,DR3}$.  
There is a trend between $\rm \Delta [Fe/H]_{phot,DR3}$ and $\rm [\alpha/Fe]$, for all the photometric metallicity relationships.  
If a linear relation is adopted, all relations show a significant slope, reported in the legend for each panel and in Table~\ref{tab:review_phot_feh}.  
In all cases, the photometric metallicities tend to under-predict $\rm [Fe/H]$ for RRLs depleted in $\rm [\alpha/Fe]$ 
and over-predict the metallicity of RRLs enhanced in $\rm [\alpha/Fe]$.  That the alpha-abundance affects 
$\rm [Fe/H]_{phot}$ is not surprising as RRLs are often enhanced in $\alpha$ but depleted in $\rm [Fe/H]$.
Therefore, it is not inconceivable that $\alpha$ plays a significant role in the shape of the RRL lightcurve, which is what $\phi_{31}$ measures. 

One of the advantages of $\rm [Fe/H]_{CaT}$ is that there does not appear to be the same kinds of 
biases with $\rm [\alpha/Fe]$ abundance that are present in metallicities derived from photometry.  
The difference between $\rm [Fe/H]_{CaT}$ and $\rm [Fe/H]_{phot}$ may also indicate the $\rm [\alpha/Fe]$ 
of a star.  RRL stars with $\rm \Delta[Fe/H] >$ 0.3 are likely to have low $\rm [\alpha/Fe]$ abundances.

We list in Table~\ref{tab:gaia_catfeh} CaT metallicities derived for the {\it Gaia} RRLs with RVS spectra.  
Only a small portion of the table is shown here for clarity and the full table is available online.

\begin{deluxetable*}{lccccccc}
\tabletypesize{\scriptsize}
\tablenum{3}
\tablecaption{The $\rm [Fe/H]$ metallicities of local RRLs from {\it Gaia} RVS spectra using the EW of the 8498\AA~Ca {\tt II} line. \label{tab:gaia_catfeh}}
\tablehead{
\colhead{{\it Gaia} ID} & \colhead{RA (deg)} & \colhead{Dec (deg)} & \colhead{EW} & \colhead{$\rm [Fe/H]$} & \colhead{S/N} & \colhead{\# Clean Epochs} & \colhead{$G$-mag }
}
\decimalcolnumbers
\startdata
1026974891482561792 & 129.357206 & 49.268806 & 512.69 $\pm$ 23.86  &-1.82 $\pm$ 0.17 & 25 &  17  & 12.068 \\
1032533743099023360 & 122.009609 & 53.658476 & 537.67 $\pm$ 23.07  &-1.73 $\pm$ 0.17 & 27  & 21 & 12.299 \\
1036978866747819264 & 132.811239 & 56.355288 & 554.42 $\pm$ 10.44  &-1.66 $\pm$ 0.15 & 60 &  25 & 11.250 \\
1050307829598059008 & 149.453252 & 60.741150 & 496.99 $\pm$ 31.49  &-1.88 $\pm$ 0.18 & 29 &  50 & 13.131 \\
1063808840251264128 & 146.736001 & 63.695459 & 659.17 $\pm$ 7.70  &-1.27 $\pm$ 0.14 & 79 &  38 & 10.883 \\
1093345639583902592 & 126.102594 & 65.717379 & 510.84 $\pm$ 7.29  &-1.83 $\pm$ 0.14 & 75  & 26 & 11.000 \\
1094115263363633408 & 127.130048 & 67.496421 & 405.13 $\pm$ 18.57  &-2.23 $\pm$ 0.16 & 29 &  43 & 12.598 \\
1104405489608579584 & 98.499256 & 67.025256 & 694.25 $\pm$ 34.98  &-1.13 $\pm$ 0.19 & 24 &  33 & 12.799 \\
1151084568571408384 & 125.681151 & 86.081024 & 490.75 $\pm$ 22.12  &-1.91 $\pm$ 0.16 & 26  & 23 & 12.302 \\
\hline
\enddata
\end{deluxetable*}


\section{Calcium Triplet Metallicities of BRAVA-RR RRLs} \label{sec:bravarr}
The Bulge Radial Velocity Assay for RR Lyrae stars (BRAVA-RR) is a spectroscopic survey targeting RRLs in the 
inner Galaxy \citep{kunder16, kunder20}.  To date $\sim$2700 RRLs have been observed using AAOmega 
on the Anglo-Australian Telescope with a similar resolution and wavelength regime as {\it Gaia}.  The BRAVA-RR 
results have shown that the RRLs in the inner Galaxy are not a homogeneous population; some follow the 
barred Bulge and some have properties that indicate they formed before the bar.  In particular, the most centrally 
concentrated RRLs have tight orbits with large eccentricities and a spatial distribution that does not show a 
clear barred distribution \citep[see Figure~6][]{kunder20}.  
Whereas this population is elusive in most inner Galaxy studies targeting the more numerous metal-rich Bulge stars 
\citep[e.g.,][]{queiroz21}, recent studies of metal-poor giants have confirmed an ancient, centrally concentrated stellar population that was part of the first few massive progenitors to form the proto-Galaxy \citep{belokurov22, arentsen24}.  

Accretion remnants will also reside in the inner Galaxy, and it is well-known that different progenitor 
populations lead to different loci in total energy vs angular momentum space \citep[e.g.,][]{massari19}.  
\citet{das20} also show that accreted structures can be identified from their Mg, Mn and Fe abundances.  
In this way, the detection of the {\it Gaia}-Enceladus/Sausage system 
\citep[GE/S,][]{belokurov18, helmi18}, the Splash accretion \citep{belokurov20}, and the inner Galaxy 
structure known as Heracles \citep{horta21}\footnote{Heracles likely arose from the Kraken event predicted by 
\citet{kruijssen20}, as they both have similar low energies, but a direct connection between Kraken and Heracles has not been established.}, have been detected within the central $\sim$4~kpc of the Galaxy using bright giant stars.  

Giants will span a wide range of ages, and this can dilute or introduce noise into the signature of old accreted 
systems.  Selecting only stars with low metallicities can help cull the younger giants, but selecting stars by 
their age is difficult.  RRLs, in contrast to giants, are exclusively an old population and are therefore useful to 
probe the first large accretion events in the Galaxy.  None of the above mentioned accretion events have been 
seen within the inner Galaxy RRL population yet, although GE/S has been detected in the local RRL 
population \citep[e.g.,][]{prudil20}.  

The lack of spectroscopic metallicities for the inner Galaxy RRLs make it difficult to chemically link any 
dynamical properties of inner Galaxy RRLs to the accretion history of the MW.  
In particular, accreted stars 
will have depleted Mg abundances ($\rm [Mg/Fe] <$0.2) for stars more metal-rich than $\rm [Fe/H] \sim -$1.25.
The $\rm [Fe/H]_{CaT}$ metallicities enable a search for stars that could be in the GE/S, Splash or IGS.

\subsection{Use of Calcium Triplet [Fe/H] abundances for Galactic Bulge RRLs}
From the BRAVA-RR Survey, the 83 RRLs with $>$8 epochs of observations were selected with which to determine $\rm [Fe/H]_{CaT}$.  The numerous epochs allow the investigation of the relationship between $\rm [Fe/H]_{CaT}$ with phase to be investigated.  
Also, multiple epochs can be averaged together for a more robust $\rm [Fe/H]_{CaT}$ metallicity, which is helpful because the S/N for the BRAVA-RR spectra is typically $\sim$10-40.

\begin{figure}
\centering
\mbox{\subfigure{\includegraphics[height=5.25cm]{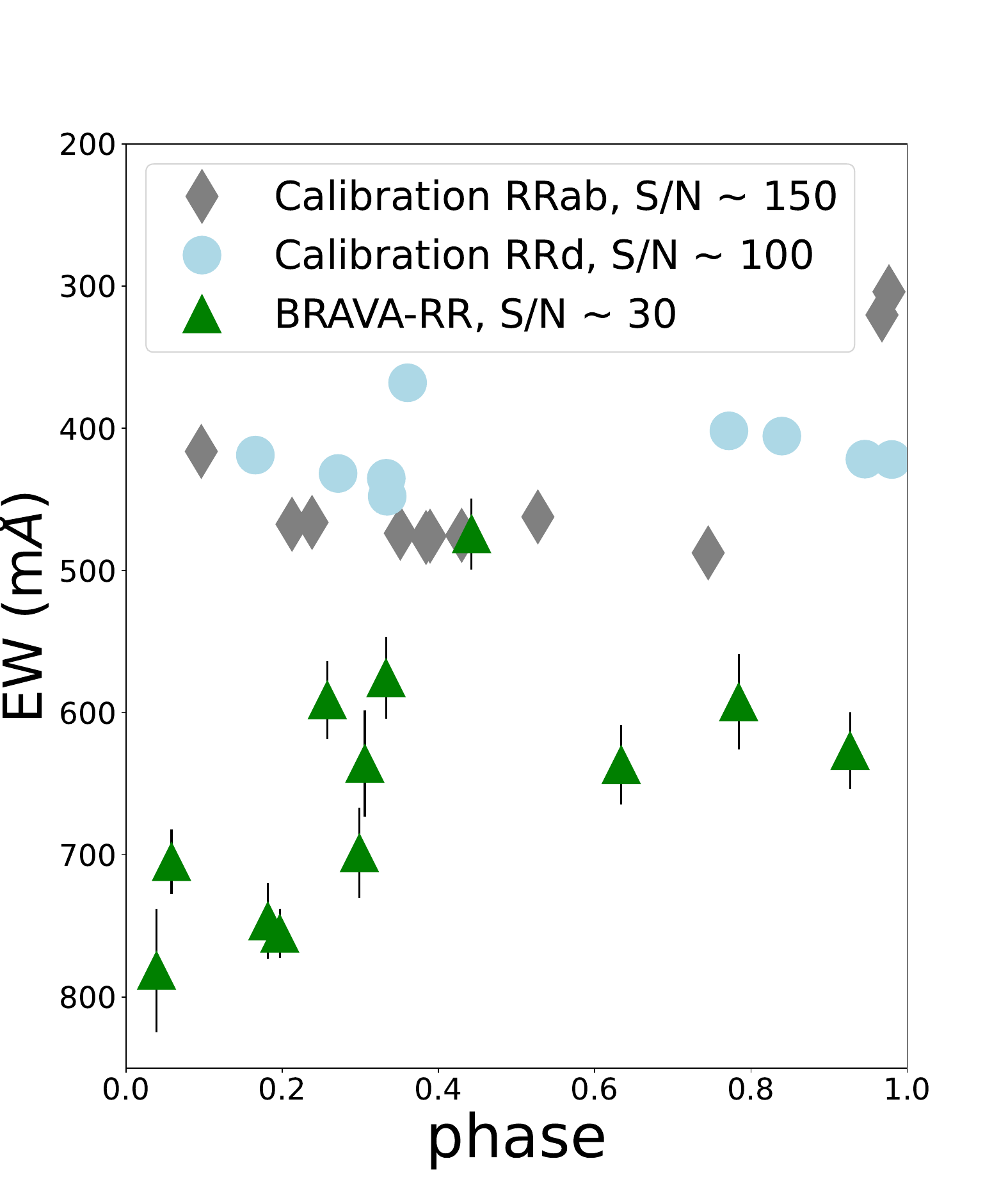}}}
{\subfigure{\includegraphics[height=5.25cm]{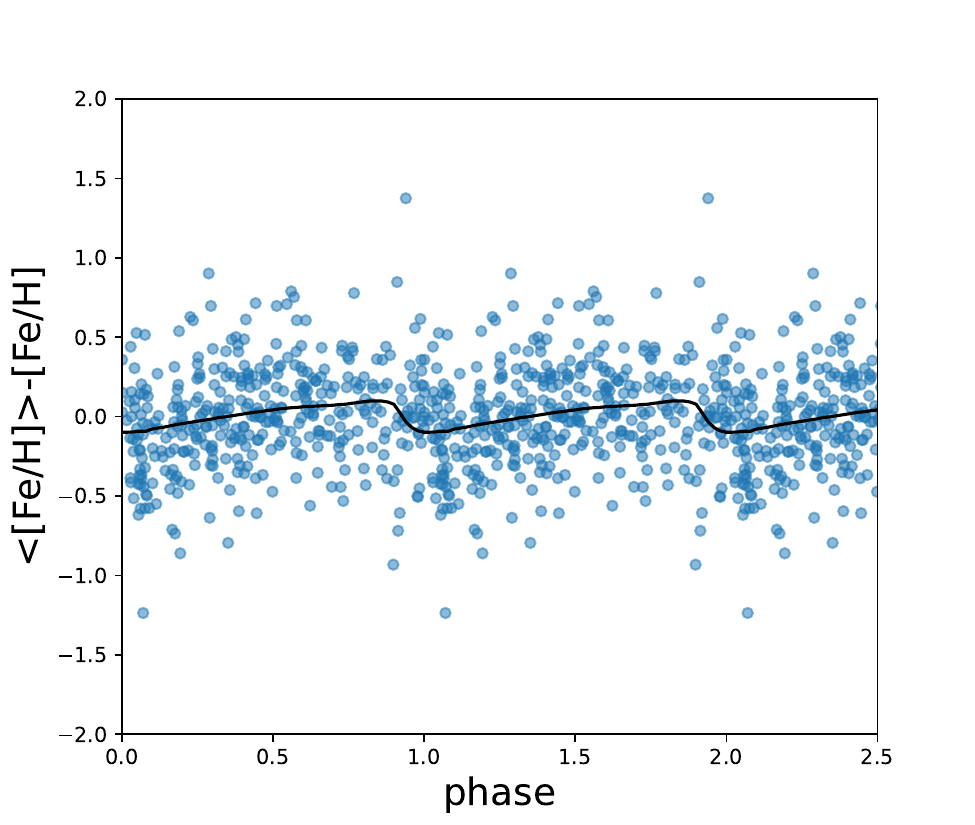}}}
{\subfigure{\includegraphics[height=5.25cm]{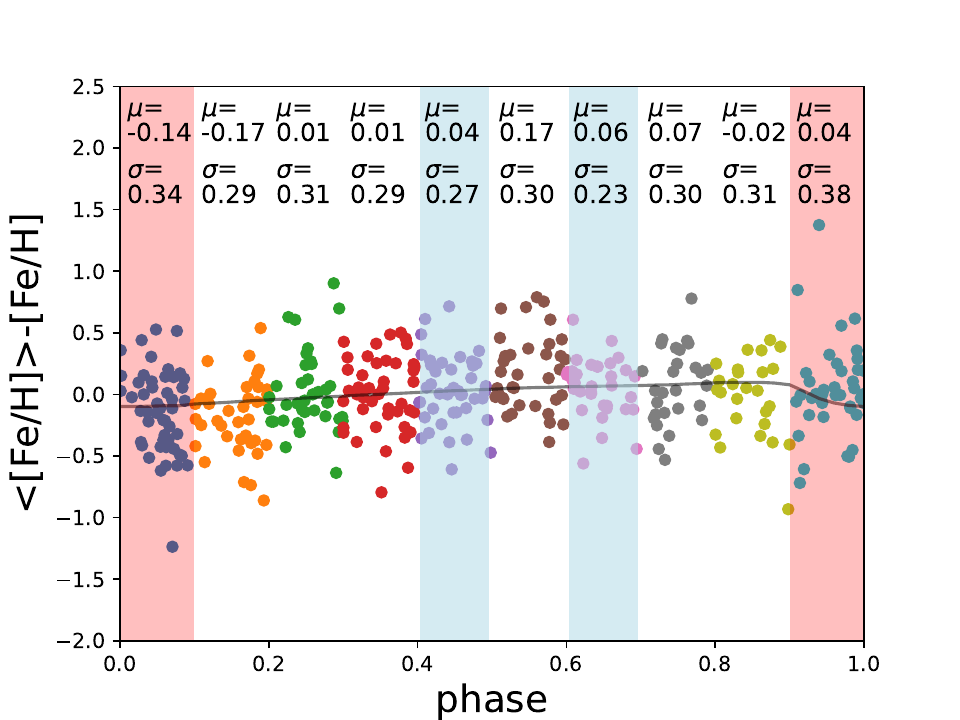}}}
\caption{
{\it Left:} An example of how the EW of a RR Lyrae stars changes with phase. {\it Middle:} The difference between each individual $\rm [Fe/H]_{CaT}$ measurement as compared to the stars $\rm <[Fe/H]_{CaT}>$.  The light-curve of a type ab RRL is overplotted. 
{\it Right:} Same as the middle panel, only the scatter about each phase is reported.  Phases with the largest scatter in $\rm [Fe/H]_{CaT}$ are highlighted in red, and the optimal phases with low $\rm [Fe/H]_{CaT}$ scatters are highlighted in blue.}
\label{bravarr_overview}
\end{figure}

The left panel in Figure~\ref{bravarr_overview} shows the measured EW as a function of phase for the star OGLE-BLG-RRLYR-10715, as well as for the two calibration stars with the highest number of epochs, the RRab star V~Ind and the RRd star V5644~Sgr (see Table~\ref{tab:CaTcalib}).  The scatter in the measured EW with phase is more prominent in the low S/N stars, and only minor scatter in EW is observed in spectra with high S/N ratios.

In order to combine the full sample of RRLs and determine how the pulsation phase affects $\rm [Fe/H]_{CaT}$, we first determine the mean $\rm [Fe/H]_{CaT}$ from all of the stars' measurements, $\rm <[Fe/H]_{CaT}>$.  Only spectra with S/N $>$25 are included in our analysis, which is 60\% of the BRAVA-RR measurements for our 83 RRL star sample.   
The middle panel in Figure~\ref{bravarr_overview} shows the difference between $\rm <[Fe/H]_{CaT}>$ and the $\rm [Fe/H]_{CaT}$ of 659 measurements for the 83 individual BRAVA-RRLs in our sample.  The characteristic saw-tooth shape of the RRL pulsations are seen, and the \citet{liu91} template is overplotted.  

The mean difference and scatter of the individual $\rm [Fe/H]_{CaT}$ measurement for each phase can be calculated, and this is shown in the right panel of Figure~\ref{bravarr_overview}.  
The mean difference in the individual $\rm [Fe/H]_{CaT}$ measurement for each phase, $\mu$, is in agreement that the systemic velocity (which can also be thought of as the zero-velocity phase, since this is the best approximation of a "static" star behavior) occurs at $\phi \sim$0.38 \citet{liu91, kunder20, prudil24b}.  

It is not surprising that phases with $\phi >$ 0.9 and $\phi <$ 0.1 have the largest scatter; 
these phases are often reported to correspond to a shocked atmosphere 
phase \citep{pancino15, kolenberg10} and also rapid changes in temperature \citep[e.g.,][]{preston22} and hence are regions where the spectra may be affected by shocks.
The phases with the smallest scatter are those between $\phi =$0.4-0.5 and $\phi =$0.6-0.7.  
Due to the somewhat low S/N of the spectra and the relatively small sample of RRLs, it may be that the smaller scatter in the difference between $\rm <[Fe/H]_{CaT}>$ and the $\rm [Fe/H]_{CaT}$ is just due to a stochastic process.  
However, there are physical indications that would give rise to the scatter being seen, for example, that the largest scatter is when the RRL is at the phase where shocks are strongest, and that the smallest scatter is at phases that have previously been observed to be a quiescent part of the RRL cycle \citep[e.g.,][]{pancino15}.  

We can obtain $\rm <[Fe/H]_{CaT}>$ using observations that fall only between those phases, 
$\rm <[Fe/H]_{CaT}>_{\phi,optimal}$.  In this case, there will be fewer $\rm [Fe/H]_{CaT}$ observations to measure 
a Calcium triplet metallicity, but the individual $\rm [Fe/H]_{CaT}$ measurements may be more accurate.  

\begin{figure}
\centering
\mbox{\subfigure{\includegraphics[height=6.7cm]{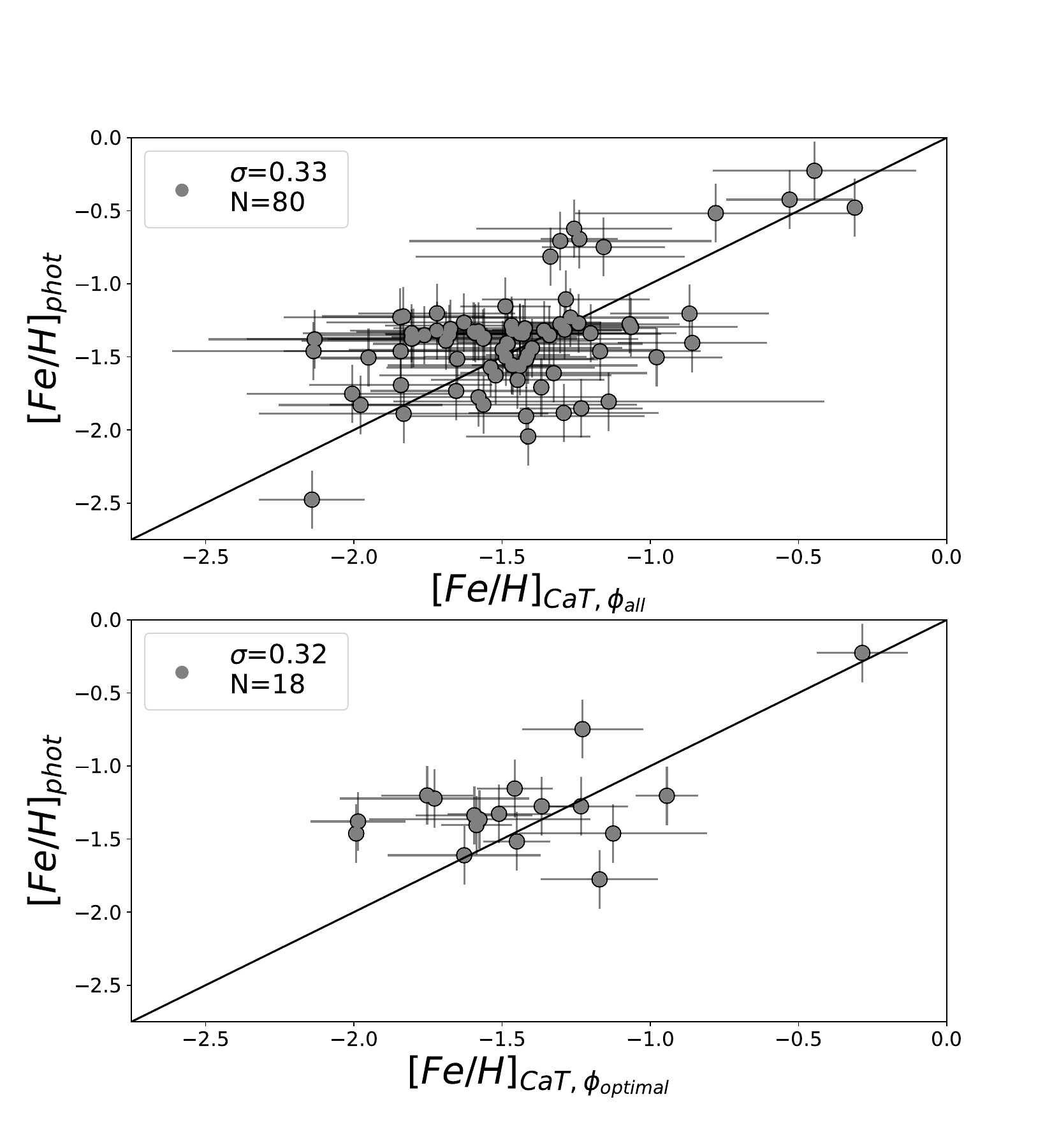}}}
{\subfigure{\includegraphics[height=6.7cm]{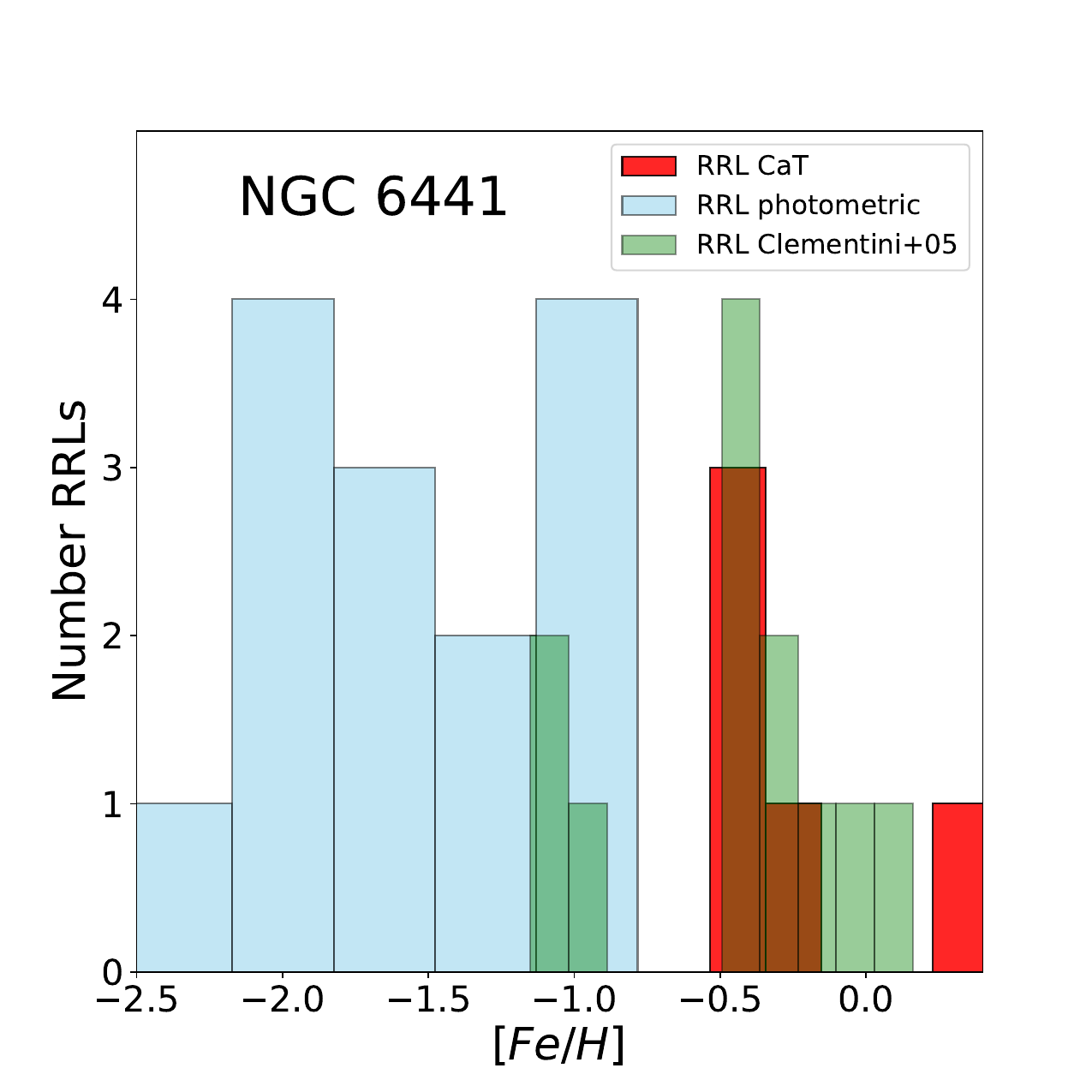}}}
\caption{
{\it Left:} $\rm [Fe/H]_{phot}$ metallicities of Galactic bulge stars as compared to our $\rm [Fe/H]_{CaT}$ metallicities, where $\rm <[Fe/H]_{CaT, \phi_{all}}>$ is the metallicity from individual $\rm [Fe/H]_{CaT}$ measurements are averaged over all phases, and $\rm [Fe/H]_{CaT, \phi_{optimal}}$ is where only $\rm [Fe/H]_{CaT}$ at the optimal phases of $\phi =$0.4-0.5 and $\phi =$0.6-0.7 are used.
{\it Right:} The $\rm [Fe/H]$ metallicity distribution of the RRLs in NGC~6441 from $\rm [Fe/H]_{CaT}$, 
$\rm [Fe/H]_{phot}$ and from \citet{clementini05}.  Unlike $\rm [Fe/H]_{photo}$, 
$\rm [Fe/H]_{CaT}$ is also able reproduce the metallicities 
of the peculiar, long-period metal-rich RRLs seen in the massive GCs NGC~6441 and NGC~6388.
}
\label{bul_cat_phot}
\end{figure}

Figure~\ref{bul_cat_phot} shows how $\rm <[Fe/H]_{CaT}>$ and $\rm <[Fe/H]_{CaT}>_{\phi,optimal}$ compare to the 
photometric metallicities, where only stars with 3 or more $\rm [Fe/H]_{CaT}$ observations are included.  
For both cases, the correlation between $\rm <[Fe/H]_{CaT}>$ and the photometric metallicities give a similar 
dispersion, agreeing with the photometric $\rm [Fe/H]$ to within $\sigma = 0.35$.  
For the case comparing $\rm <[Fe/H]_{CaT}>_{\phi,optimal}$ with the photometric metallicities, the bulk of the RRLs have $\rm [Fe/H] \sim$ $-$1.5, and the slope is dominated by one metal-rich RRL.

Of particular interest is how well $\rm [Fe/H]_{CaT}$ metallicities can reproduce the metal-rich RRLs with 
long periods; such RRLs have to date only been observed in the Galactic bulge, mainly in the massive 
GCs NGC~6441 and NGC~6388.  Photometric metallicities notoriously fail to reproduce the 
high $\rm [Fe/H]$ metallicities for such long period RRLs \citep{clementini05}.  
NGC~6441 and NGC~6388 have extended HBs and significant numbers of RRLs 
with very long periods of $\sim$ 0.75 days, but are metal-rich \citep{pritzl00, pritzl01}.  Metal-rich GCs tend to have 
very few or no RRLs, and metal-rich RRLs typically have shorter periods compared to their metal-poor counterparts, 
so the properties of the RRLs in those GCs are surprising and no satisfactory explanation for the cluster’s peculiar RRL population exists.  

None of our 83 RRLs are part of the GCs NGC~6441 or NGC~6388, but a handful of stars in the full 
BRAVA-RR sample do belong to NGC~6441.  Five of those stars were observed at optimal phases, and in this way, 
$\rm <[Fe/H]_{CaT}>_{\phi,optimal}$ can be calculated.  
Figure~\ref{bul_cat_phot} shows the $\rm [Fe/H]_{phot}$ as compared to the medium-resolution 
spectroscopic metallicities of \citet{clementini05} and as compared to our 
$\rm [Fe/H]_{CaT, \phi_{optimal}}$ metallicities for the Globular Cluster NGC~6441.    
Our sample size is small, but the $\rm [Fe/H]_{CaT}$ metallicities are able to recover the metal-rich 
nature of the NGC~6441 RRLs, in contrast to $\rm [Fe/H]_{phot}$.

We adopt the average $\rm [Fe/H]_{CaT}$ metallicities of all measurements, $\rm <[Fe/H]_{CaT}>$, as the metallicity for our Galactic bulge RRL sample.  
There are 80 RRLs that have between 3 and 12 epochs of observations with S/N $>$25.  
In a future paper, we will expand our analysis to also include those BRAVA-RR stars with $<$8 epochs of observations that happen to have observations taken at optimal phases (such as we did with the RRLs in NGC~6441).  

The average metallicities of $\rm <[Fe/H]_{phot}>$ and $\rm <[Fe/H]_{CaT}>$ of the sample are almost identical at $\rm [Fe/H] \sim -$1.4, with also identical dispersions of $\sigma = 0.36$.  This is consistent with what is typically reported in the literature: that photometric metallicities are in agreement with the average metallicity of a population of RRLs, but individual measurements of RRLs may not always be accurate \citep[e.g.,][]{clementini23, dekany21}.  

\subsection{Chemodynamics of Galactic Bulge RRLs}

To estimate the orbital properties, we used the \texttt{galpy}\footnote{\url{https://docs.galpy.org/en/v1.9.2/}} module \citep{bovy15} and implemented the \texttt{McMillan17} potential \citep{mcmillan17} in the aforementioned package. 
For the Sun’s Galactocentric location and rotation velocity, we adopted $R_{0} = 8.21$\,kpc and $v_{0} = 233.1$\,km\,s$^{-1}$, as specified in the \texttt{McMillan17} potential. 
We utilized values estimated by \citet{schonrich2010}, $\left ( U_{\odot}, V_{\odot}, W_{\odot} \right ) = \left ( 11.1, 12.24, 7.25 \right )$. 
For each RRL, we varied its observed properties within their errors during orbital integration ($t=2$\,Gyr) and estimated the average orbital properties (e.g., angular momentum $L_z$ and orbital energy $E_{\mathrm{tot}}$) along with their dispersion.

\begin{figure}
\centering
\mbox{\subfigure{\includegraphics[height=5.0cm]{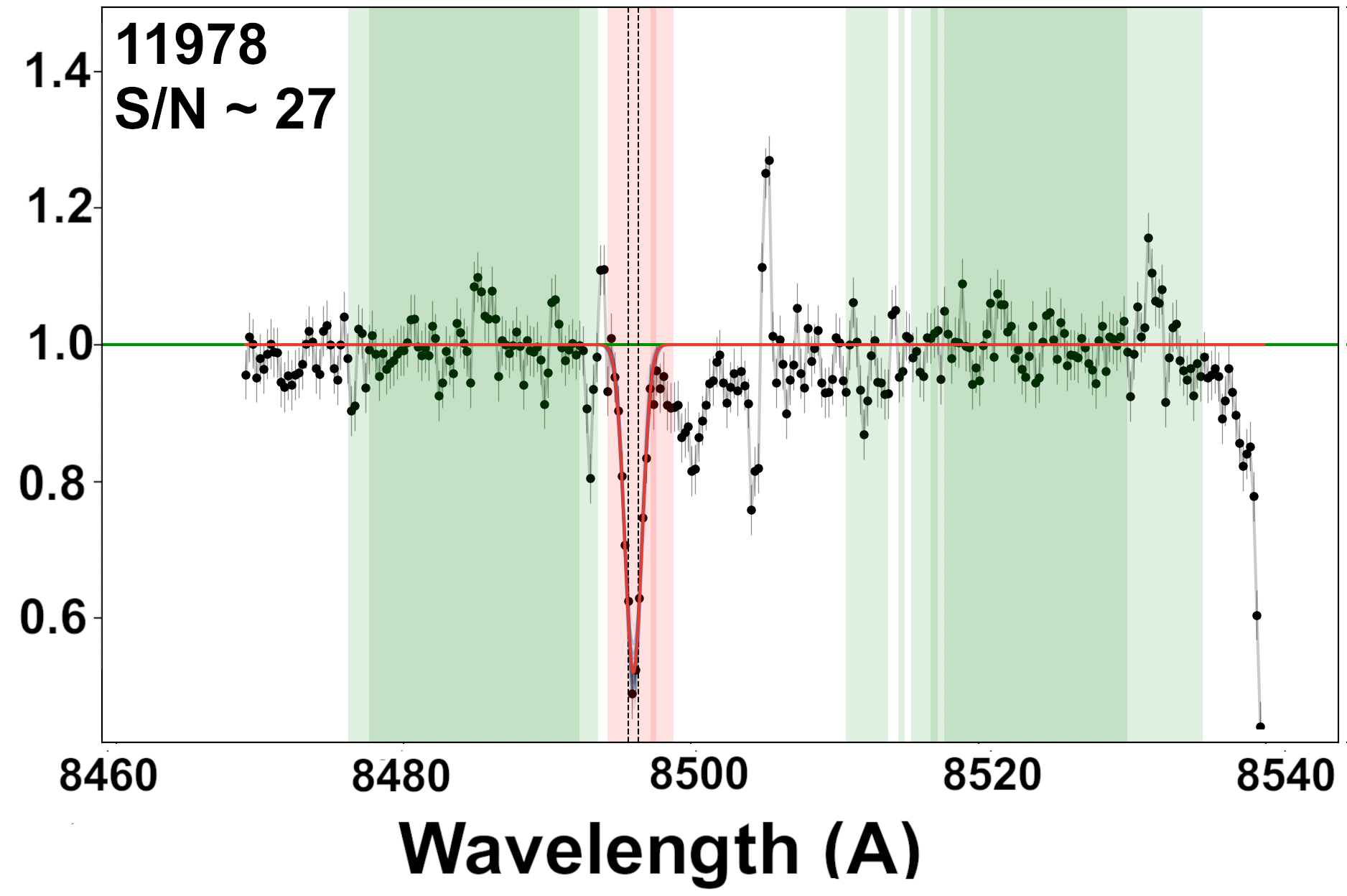}}}
{\subfigure{\includegraphics[height=6.7cm]{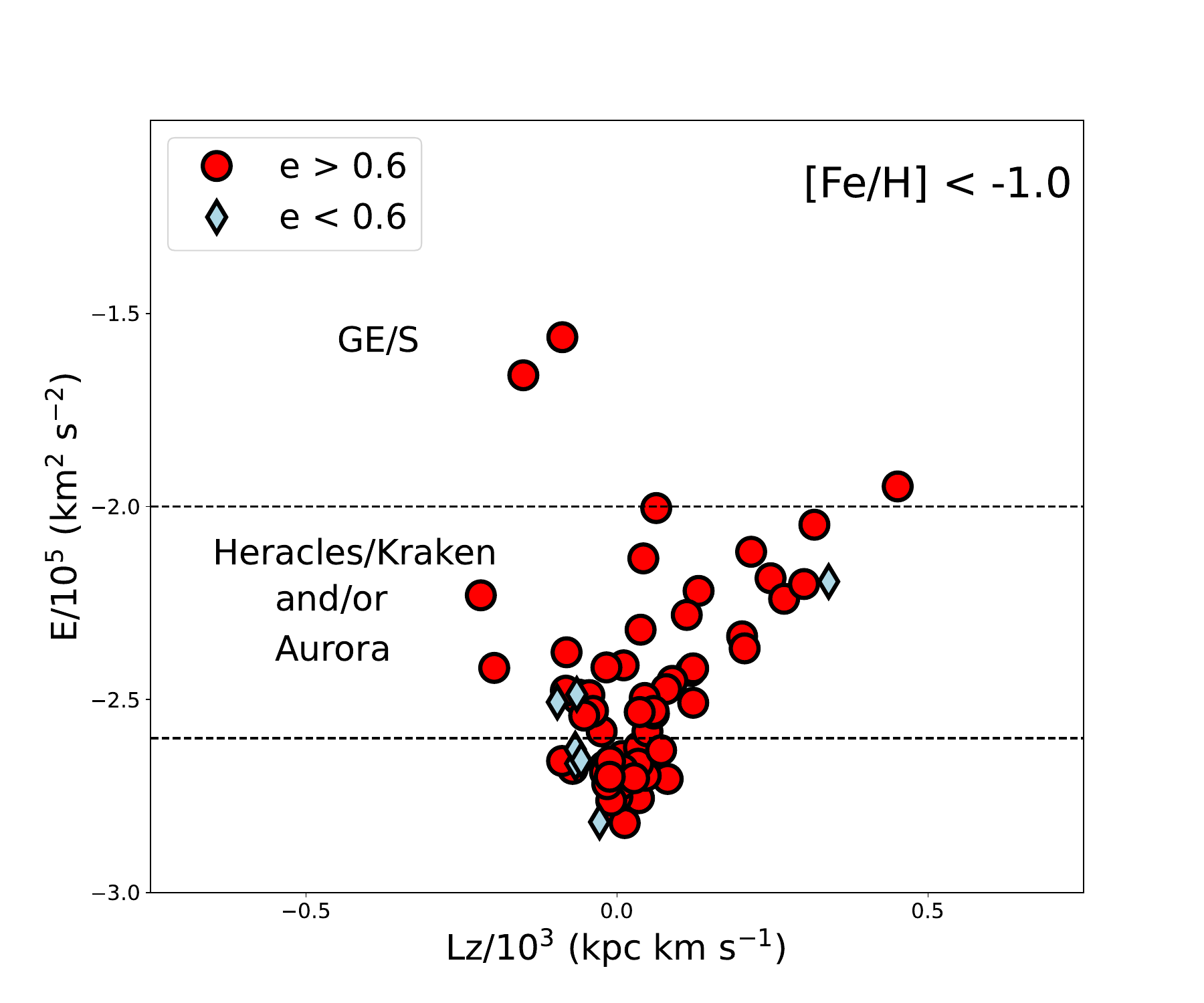}}}
\caption{
{\it Left:} A typical BRAVA-RR spectra used for $\rm [Fe/H]_{CaT}$ abundance determination from the measured EW of the CaT line at 8498\AA. {\it Right:} The total energy ($E_{tot}$) versus the $z$ component of the angular momentum ($L_z$) for the Galactic bulge RRLs 
studied here.  Stars belonging to the same physical structure cluster together in integrals of motion such as 
$E_{tot}$ and angular momentum, and the major substructure of GE/S is labelled.  The $E_{tot}$ and angular momentum regime where Aurora is seen is also where the potential Heracles/Kraken substructure is thought to resided.
}
\label{figIGS}
\end{figure}

Figure~\ref{figIGS} (right panel) shows the ($L_z$, $E$) orbit distribution of the RRLs with 
metallicities more metal-poor than $\rm [Fe/H]_{CaT} <$ $-$1.0.  The stellar orbit distribution depends 
mainly on the metallicity and radial distance of the stars probed.  
Probing small radial distances from the Galactic center gives the possibility to detect stars with the lowest energies.   
The Galactic disk and the $in situ$ populations will dominate at $\rm [Fe/H] > -$1.0, and the halo and possible accreted populations will become apparent at $\rm [Fe/H] <-$1.0.  
Simulations of MW-like galaxies indicate that mergers contribute to a small fraction of the stellar component, and so the majority of stars with low metallicities and energies will belong to the $in situ$ component of the MW \citep[e.g.,][]{gargiulo19, semenov23}.
In an attempt to uncover accreted structures, Mg, Mn, Al and Fe abundances \citep[e.g.,][]{hawkins15,das20, naidu20} are often used.  
Using these elemental abundances, primarily $\rm [Al/Fe]$, it has also been put forward that almost all stars in the inner Galaxy 
with $\rm [Fe/H] < -$1.0 and low energies formed {\it in situ} in what has been dubbed {\it Aurora} \citep{belokurov22}.  
This is because stars at low metallicities and low energies 
consistently have higher $\rm [Al/Fe]$ ratios as compared to the same stars at higher energies, 
whereas accreted stars would be depleted in $\rm [Al/Fe]$.  
However, using similar diagnostics, it was also been put forward that stars at low energies and metallicities can belong to a massive system that merged early in the history of the Milky Way, the Kraken/Heracles \citep{kruijssen20, horta21}.  
It is still not clear if Heracles/Kraken is in fact a separate accreted component in the Milky Way or if Aurora contains most or all of the alleged Kraken/Heracles structure.  

In Figure~\ref{figIGS}, two RRLs are consistent with belonging to the low-metallicity 
sub-structure discussed in the introduction, GE/S.  
A striking 58 per cent of our sample consist of RRLs consistent with Heracles/Kraken, as they have highly eccentric orbits and with energies between $-$2x10$^5$ km$^2$~s$^{-2}$ and $-$2.6x10$^5$ km$^2$~s$^{-2}$.  
Due to the $\rm [Fe/H]_{CaT}$ uncertainties, there may be a few metal-rich contaminants that ended up in this sample, but given the paucity of metal-rich RRLs (see $e.g.$, left panel of Figure~\ref{bul_cat_phot}), this should be a small number and would not affect our conclusions.  
This energy regime also overlaps with the $in situ$ Aurora regime.  
The high fraction of RRLs found with these energies is not compatible with these RRLs belonging be an accreted component alone -- accreted components will not dominate in the inner Galaxy -- and therefore we interpret the low-energy RRLs as belonging mainly to the $in situ$ Aurora bulge.  It may be that that there is some contamination from accretion,  but with the absence of elemental abundances for these RRLs, it is difficult to make firm conclusions.

\section{Conclusions}
Here 68 spectra from a sample of 40 RRLs with metallicities derived from Fe(II) and Fe(I) abundances are used to verify the correlation between $\rm [Fe/H]$ and the EW of the Ca {\tt II} line at 8498 \AA~to for RRLs.  
The correlation we obtain is almost identical to that obtained by \citet{wallerstein12}, but our relation has associated uncertainties, which allows error estimations on our derived $\rm [Fe/H]_{CaT}$ metallicities.  
Our $\rm [Fe/H]_{CaT}$ metallicities are a factor of 2 more precise than 
the {\it Gaia} photometric metallicities released in DR3 \citep{clementini23}, and our metallicities are especially 
accurate for spectra with S/N $>$ 35.  We publicly release 175 spectroscopic metallicities for the {\it Gaia} RRLs 
with RVS spectra available to the community.   This number will increase dramatically for the next {\it Gaia} releases.  
Our analysis suggests that the average $\rm [Fe/H]_{CaT}$ measurement, $\rm <[Fe/H]_{CaT}>$, is a good reference for the metallicity of an RRL, and this will especially hold when the spectra evenly cover the whole pulsation cycle.  
Given the {\it Gaia} scanning law, this will be the case for most of the {\it Gaia} RRL sample.  

We show that photometric metallicities are influenced by their $\rm [\alpha/Fe]$, in a sense that $\rm [Fe/H]_{phot}$ 
appears to be more depleted in $\rm [Fe/H]$ for RRLs with lower $\rm [\alpha/Fe]$ ratios.  This bias is not observed, or at least is not as strong, for metallicities derived from the EW of CaT.  We further show that $\rm [Fe/H]_{CaT}$ correctly 
recovers the high $\rm [Fe/H]$ abundances observed in the long-period metal-rich population of RRLs in NGC~6441.  
This has been a challenge for photometric metallicity relationships.

Using a sample of 80 RRLs with $>$8 epochs of observations in the BRAVA-RR survey, we find a small number of RRLs with properties consistent with belonging to the accreted GE/S substructure.  We find a much larger number of RRLs consistent with having high-eccentricities, low-energies and without a prominent rotating component, which are properties of Kraken/Heracles.  Due to the large fraction ($>$50\%) of the RRL population inhabiting such a low-energy regime, these stars were likely not accreted, but are part of the $in situ$ Milky Way stellar population that formed before the Galaxy had a coherently rotating disc, Aurora.
The commanding presence of RRLs with properties 
consistent of Aurora, much more so than $e.g.,$ the RRLs with properties of the GE/S, is in agreement with Aurora being formed sufficiently early, before any coherent disk is established. 

The relationship between the EW of the CaT and $\rm [Fe/H]$ allows one of the fundamental properties, 
$\rm [Fe/H]$, of RRLs to be derived for stars at the distance of the Galactic bulge 
and beyond, as well as for the thousands of RRLs observed with {\it Gaia} RVS.  We will apply our relationship 
to a wider sample of BRAVA-RR stars, also to those with $<$8 epochs, as well as to {\it Gaia} RRLs observed with RVS and released in upcoming DR4.

\section*{Acknowledgments}
AMK acknowledges support from grant AST-2009836 from the National Science Foundation.  
AMK, CS, KRC, JH, KD acknowledge the M.J. Murdock Charitable Trust's support through its RAISE (Research Across Institutions for Scientific Empowerment) program.  
This work was made possible through the Preparing for
Astrophysics with LSST Program, supported by the Heising-Simons Foundation and
managed by Las Cumbres Observatory.  
This research was supported by the Munich Institute for Astro-, Particle and BioPhysics (MIAPbP) which is funded by the Deutsche Forschungsgemeinschaft (DFG, German Research Foundation) under Germany´s Excellence Strategy – EXC-2094 – 390783311.

This work has made use of data from the European Space Agency (ESA) mission {\it Gaia} (\url{https://www.cosmos.esa.int/gaia}), processed by the {\it Gaia} Data Processing and Analysis Consortium (DPAC, \url{https://www.cosmos.esa.int/web/gaia/dpac/consortium}). Funding for the DPAC has been provided by national institutions, in particular the institutions participating in the {\it Gaia} Multilateral Agreement.


\clearpage


\rotate
\begin{deluxetable*}{ccccccccccc}
\tabletypesize{\scriptsize}
\tablenum{1}
\tablecaption{RR Lyrae star spectra used to calibrate the EW for Ca \texttt{II} at $8498$\,\AA~ to $\rm [Fe/H]$ metallicity
\label{tab:CaTcalib}}
\tablehead{
\colhead{GaiaID} & \colhead{$\rm [Fe/H]$} & \colhead{Per} & \colhead{$M_0$} & \colhead{HJD}  & \colhead{Type} & \colhead{name} & \colhead{S/N} & \colhead{resolution} & \colhead{instrument} & \colhead{EW} 
}
\decimalcolnumbers
\startdata
\hline
6539269990871163392 & $-$1.78 $\pm$ 0.11 & 0.6587047 & 2453694.233837 & 2457288.627074 & RRab & ADP.2017-05-12T07:46:45.719 & 59  & 11333  & XSHOOTER   & 577.1 $\pm$ 19.8 \\
6677541412082758144 & $-$2.39 $\pm$ 0.11 & 0.4069942 & 2454330.624998 & 2457288.547716 & RRc  & ADP.2017-05-12T07:46:45.749 & 41  & 11333  & XSHOOTER   & 419.6 $\pm$ 12.5 \\
6483680332235888896 & $-$1.62 $\pm$ 0.01 & 0.4795935 & 2457223.412508 & 2457618.643778 & RRab & ADP.2017-05-12T15:04:34.079 & 163 & 18340 & XSHOOTER   &  638.5 $\pm$ 3.8  \\
4918030715504071296 & $-$1.68 $\pm$ 0.14 & 0.3332324 & 2456833.506586 & 2454050.972451 & RRc  & 20061111\_0005m59\_080 & 36 & 7500  & RAVE       & 592.2 $\pm$ 22.9 \\
3806559592377384704 & $-$1.64 $\pm$ 0.14 & 0.4631965 & 2457128.057598 & 2457782.714295 & RRab & ADP.2017-05-12T19:12:51.203 & 51  & 8935 & XSHOOTER   & 541.7 $\pm$ 10.5 \\
2608819623000543744 & $-$2.31 $\pm$ 0.04 & 0.608874 & 2457593.585532 & 2457234.761289 & RRab & ADP.2020-06-10T17:26:03.411 & 68 & 42310 & UVES & 524.9 $\pm$ 3.0 \\
\enddata
\end{deluxetable*}

{}

\appendix

\section{Photometric metallicities}
\label{sec:appendixA}
\begin{figure}
\centering
\mbox{\subfigure{\includegraphics[height=9cm]{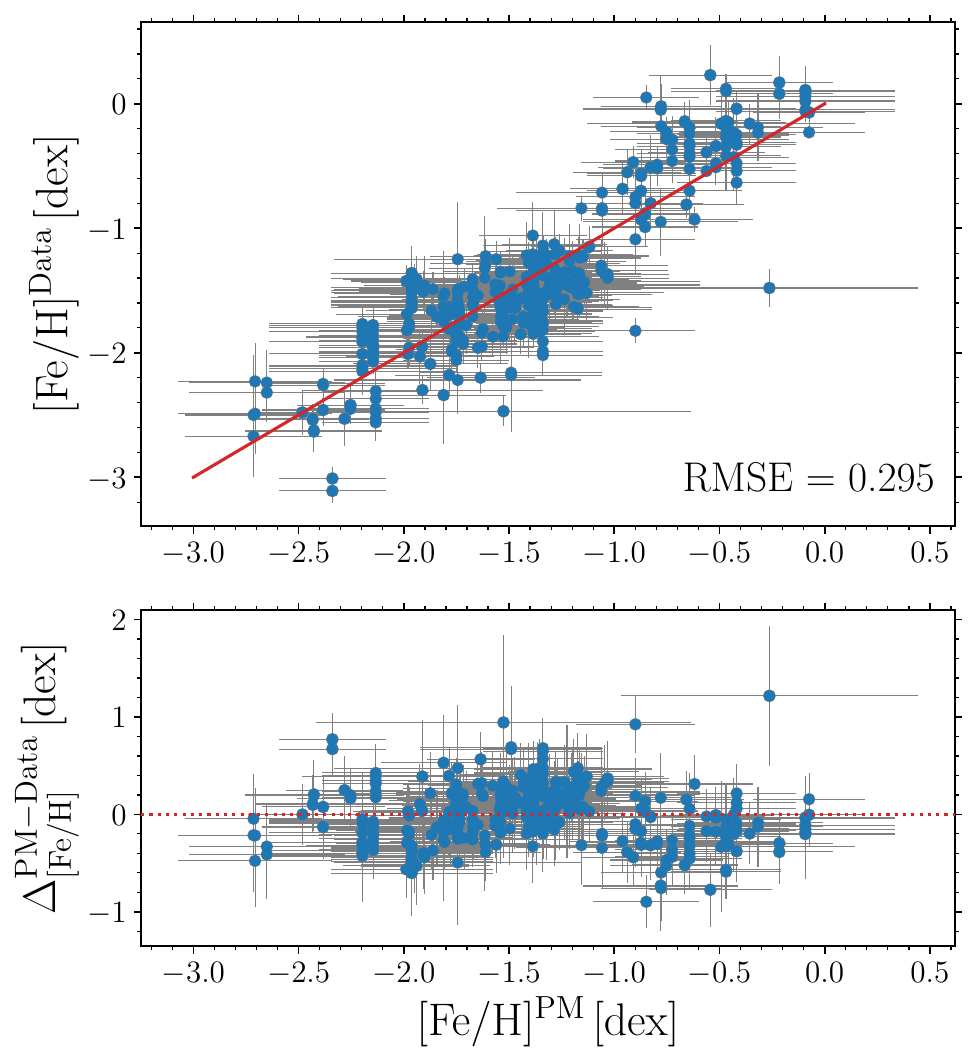}}}
\mbox{\subfigure{\includegraphics[height=9cm]{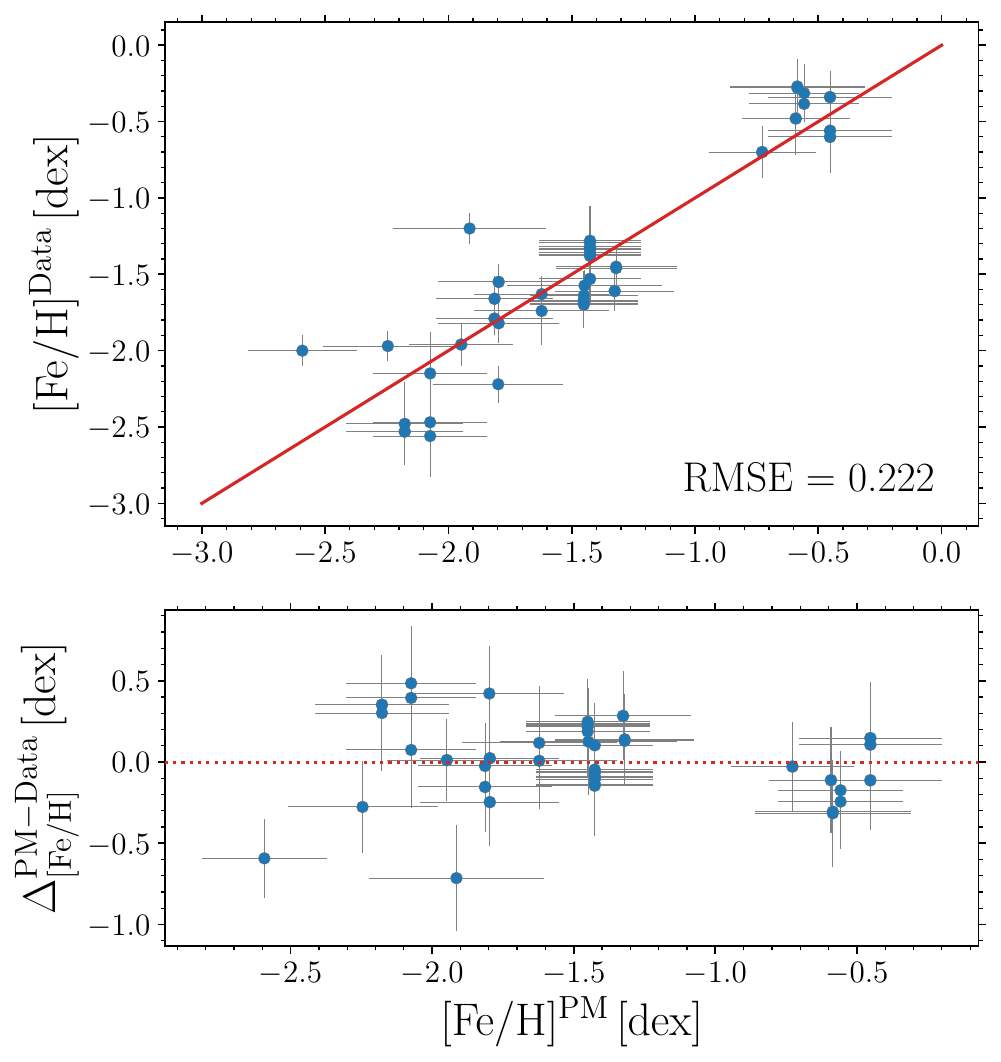}}}
\caption{
Comparison of the photometric metallicities predicted from newly derived PM relations for RRab (left-hand panel) and RRc (right-hand panel) in this study based on \textit{Gaia} $G$-band photometry.
}
\label{figPM}
\end{figure}

A number of different photometric metallicity relations based on the {\it Gaia} $G$-band photometry have been put forward.  To date, none of these use the \citet{crestani21a} sample of RRLs as a calibrating sample.  We believe this sample is superior in terms of its number of RRLs, wide metallicity baseline, high spectral resolution and homogeneity of the analysis, and here we recalibrate period-metallicity (PM) relations to estimate photometric metallicities using the publicly available photometric data provided in the \textit{Gaia} RR~Lyrae catalog \citep{clementini23}. Our approach is similar to our previous study in which we calibrate absolute magnitudes of RRLs from the {\it Gaia} RRL parallaxes \citep[see Sections 2 and 3 in][]{prudil24a}.  Briefly, spectroscopic metallicities are collected from \citet{crestani21a,crestani21b,dekany21} and matched with the corresponding star from the \textit{Gaia} RR~Lyrae catalog.  Our dataset is then divided into two groups based on the pulsation mode and stars not fulfilling the following criteria are discarded:
\begin{equation} \label{eq:Condition1}
\text{RUWE} < 1.4 \quad \text{and} \quad \texttt{ipd\_frac\_multi\_peak} < 5.
\end{equation}
This results in $20$ RRc and $155$ RRab calibration variables. In our fitting procedure, we optimize the following expression for predicting photometric metallicities:
\begin{equation} \label{eq:PM_to_calib}
\text{[Fe/H]}_{\text{phot}} = \alpha P + \beta \varphi_{31} + \gamma,
\end{equation}
where $P$ and $\varphi_{31}$ represent pulsation periods and phase differences between two Fourier coefficients, respectively, and $\alpha$, $\beta$, and $\gamma$ stand for parameters of the fit. As an additional parameter characterizing the intrinsic scatter in the photometric metallicity relation, we used $\varepsilon_{\text{[Fe/H]}}$. In the optimization procedure, we follow the described procedure in \citet{prudil24}, but we minimized the difference between spectroscopic metallicities from the literature and those predicted from Equation \ref{eq:PM_to_calib}. The results of our calibration are depicted in Figure~\ref{figPM} and summarized by the following equations, where $N$ denotes the total number of RR Lyrae variables used in calibration:
\begin{align}
\begin{split}\label{eq:1}
\text{[Fe/H]$_{\rm phot}^{\rm RRab}$}= -5.5061\,P + 0.8143\,\varphi_{31} -0.0813 , \hspace{0.05cm} N = 155
\end{split}\\
\begin{split}\label{eq:2}
\text{[Fe/H]$_{\rm phot}^{\rm RRc}$}= -13.2023\,P + 0.5138\,\varphi_{31} + 0.7338 , \hspace{0.05cm} N = 20 \hspace{2cm}.
\end{split}
\end{align}
The covariance matrices for both relations with associated intrinsic scatters are the following:
\begin{equation}
\text{Cov}_{K_{\rm s}} = 
\begin{bmatrix}
\sigma_{\alpha_{\rm RRab}} & \sigma_{\beta_{\rm RRab}} & \sigma_{\gamma_{\rm RRab}} \\
\hline
 0.019144 & -0.001096 & -0.008282 \\
-0.001096 & 0.000928 & -0.001310 \\
-0.008282 & -0.001310 & 0.007427 \\
\end{bmatrix} \\ \hspace{0.3cm} \varepsilon_{\text{[Fe/H]}_{\rm RRab}} = 0.247
\end{equation}
\begin{equation}
\text{Cov}_{\rm RRc} = 
\begin{bmatrix}
\sigma_{\alpha_{\rm RRc}} & \sigma_{\beta_{\rm RRc}} & \sigma_{\gamma_{\rm RRc}} \\
\hline
 0.758431 & -0.011111 & -0.185556 \\
-0.011111 & 0.003445 & -0.008078 \\
-0.185556 & -0.008078 & 0.082015 \\
\end{bmatrix} \\ \hspace{0.3cm} \varepsilon_{\text{[Fe/H]}_{\rm RRc}} = 0.193
\end{equation}

Figure~\ref{figPM} shows the best-fit relations.  The metallicity interval has a sufficient number of metal-rich ($\rm [Fe/H] > -0.5$) and metal-poor ($\rm [Fe/H] < -2.0$) RRLs, especially for the RRab stars (left panel).  A moderate systematic trend of the residuals for the metal-rich end is seen, where our photometric metallicity relation tends to underestimate the metallcity.  One reason for this is that the $\rm [\alpha/Fe]$ of an RRL also influences the shape of its lightcurve, and metal-rich RRLs tend to also be depleted in  $\rm [\alpha/Fe]$, unlike the 
more metal-poor RRLs.

\end{document}